%% file: sample-authordraft.tex
\documentclass[sigconf]{acmart}
\makeatletter
\def\@ACM@checkaffil{
    \if@ACM@instpresent\else
    \ClassWarningNoLine{\@classname}{No institution present for an affiliation}%
    \fi
    \if@ACM@citypresent\else
    \ClassWarningNoLine{\@classname}{No city present for an affiliation}%
    \fi
    \if@ACM@countrypresent\else
        \ClassWarningNoLine{\@classname}{No country present for an affiliation}%
    \fi
}
\makeatother

\AtBeginDocument{%
  \providecommand\BibTeX{{%
    \normalfont B\kern-0.5em{\scshape i\kern-0.25em b}\kern-0.8em\TeX}}}

\acmPrice{15.00}
\acmISBN{978-1-4503-XXXX-X/18/06}

\copyrightyear{2023}
\acmYear{2023}
\setcopyright{acmlicensed}\acmConference[KDD '23]{Proceedings of the 29th
ACM SIGKDD Conference on Knowledge Discovery and Data Mining}{August
6--10, 2023}{Long Beach, CA, USA}
\acmBooktitle{Proceedings of the 29th ACM SIGKDD Conference on Knowledge
Discovery and Data Mining (KDD '23), August 6--10, 2023, Long Beach, CA,
USA}
\acmPrice{15.00}
\acmDOI{10.1145/3580305.3599847}
\acmISBN{979-8-4007-0103-0/23/08}

\usepackage{balance}

\usepackage[final]{changes}

\usepackage{lipsum}
\definechangesauthor[name={y}, color=orange]{y}
\definechangesauthor[name={b}, color=red]{b}

\usepackage{subcaption}
\usepackage{multirow}
\usepackage{bm} 
\usepackage{amsmath,amsfonts}
\usepackage{algorithmic}
\usepackage{graphicx}
\usepackage{textcomp}
\usepackage{xcolor}
\usepackage{hyperref}
\usepackage[linesnumbered,ruled,vlined]{algorithm2e}

\SetCommentSty{mycommfont}
\SetKwInput{KwInput}{Input}                
\SetKwInput{KwOutput}{Output}              
\SetKwInput{Parameters}{Parameters}              

\graphicspath{{./img/}}

\newcommand\eat[1]{}

\newcommand\bigeat[1]{}

\begin{document}

\title{Influence Maximization with Fairness at Scale}


\author{Yuting Feng}
\affiliation{%
  \institution{University of Paris-Saclay \& CNRS LISN}
  }
\email{yuting.feng@universite-paris-saclay.fr}

\author{Ankitkumar Patel}
\affiliation{%
  \institution{University of California - Berkeley}
}
\email{ankitkumar@berkeley.edu}

\author{Bogdan Cautis}
\affiliation{%
  \institution{University of Paris-Saclay \& CNRS IPAL Singapore}
  }
\email{bogdan.cautis@universite-paris-saclay.fr}

\author{Hossein Vahabi}
\affiliation{%
  \institution{University of California - Berkeley}
  }
\email{puyavahabi@berkeley.edu}




\begin{abstract}
In this paper, we revisit the problem of \emph{influence maximization with fairness}, which aims to select $k$ influential nodes to maximise the spread of information in a network, while ensuring that selected sensitive user attributes (e.g., \emph{gender}, \emph{location}, \emph{origin, race}, etc) are fairly affected, i.e., are proportionally similar between the original network and the affected users. Recent studies on this problem focused only on extremely small networks, hence the challenge remains on how to achieve a scalable solution,  applicable to networks with millions or billions of nodes. We propose an approach that is based on learning node representations (embeddings) for fair spread from  \emph{diffusion cascades}, instead of the social connectivity, and in this way we can deal with very large graphs. We propose two data-driven approaches: (a) fairness-based participant sampling (FPS), and (b) fairness as context (FAC). Spread related user features, such as the probability of diffusing information to others, are derived from the historical information cascades, using a deep neural network. The extracted features are then used in selecting influencers that maximize the influence spread, while being also fair with respect to the chosen sensitive attributes.  In FPS, fairness and cascade length information are considered independently in the decision-making process, while FAC considers these information facets jointly and takes into account correlations between them. The proposed algorithms are generic and represent the first policy-driven solutions that can be applied to arbitrary sets of sensitive attributes at scale. We evaluate the performance of our solutions on a real-world public dataset (Sina Weibo) and on a hybrid real-synthetic dataset (Digg), which exhibit all the facets that we exploit, namely diffusion network, diffusion traces, and user profiles. These experiments show that our methods outperform the state-the-art solutions in terms of spread, fairness, and scalability.  
\end{abstract}





\maketitle

\section{Introduction}
\input{introduction.tex}

\section{Related work}
\input{related.tex}

\section{Preliminaries}
\input{preliminaries.tex}

\section{Problem Definition}
\input{problemdef.tex}

\section{Algorithmic Solutions}
\input{solution.tex}

\section{Experiments} 
\input{experiments.tex}

\section{Conclusion}
\input{conclusion.tex}

\section*{Acknowledgements}
DesCartes: this research is supported by the National Research Foundation, Prime Minister’s Office, Singapore under its Campus for Research Excellence and Technological Enterprise (CREATE) program.  This research is also supported by the CNRS grant FairIM@Scale. We would like to thank S. Skowronski, D. Peletz, A. Stover, A. Rains, T. Singh, C. S. Bikkanu, and A. Todeschini for support on code improvements and early stages of this work. 



\bibliographystyle{ACM-Reference-Format}
\balance
\bibliography{bibliography}


\appendix

\clearpage
\section{SUPPLEMENTARY MATERIAL}

\subsection{Complexity of the fair-greedy algorithm}
\label{sec:complexity}
Similar to \cite{multitask_2020},  the fair-greedy influencer selection algorithm computes the expected number of influenced nodes $\lambda[u]$ using the norm-2 of the embedding vectors. The computation of the norm-2 for a  vector of size $|V|$ requires $O(|V|)$ steps. For $|I|$ cascade initiators, the complexity of finding the expected number of influenced nodes is therefore $O(|I||V|)$. Next, the algorithm needs to find a number of $\lambda[u]$ yet-to-be-influenced nodes maximizing the cumulative spread probability, which requires sorting $O(|V|)$ nodes and picking the top $\lambda[u]$ ones, with overall $O(|V| \log  |V|)$ complexity. The sorting operation is performed for each influencer, and thus the complexity of updating the marginal gain vector  is $O(|I| |V| \log|V|)$. Finally, the influencer maximizing the marginal gain is selected, which has $O(|I|\log |I|)$ complexity. These operations are repeated until $k$ influences are found, and thus, the overall complexity becomes $O(k  |I| |V| \log |V|)$. As with CELF, the algorithm benefits from (i) the fact that $|V|$ diminishes at every iteration, and (ii) much fewer influence spread evaluations than $I$ are done (lines  $19-23$). 

\subsection{Distribution of the sensitive attributes}
\label{sec:distributions}
\begin{figure}[bth]
  \centering
  \includegraphics[width=0.8\linewidth]{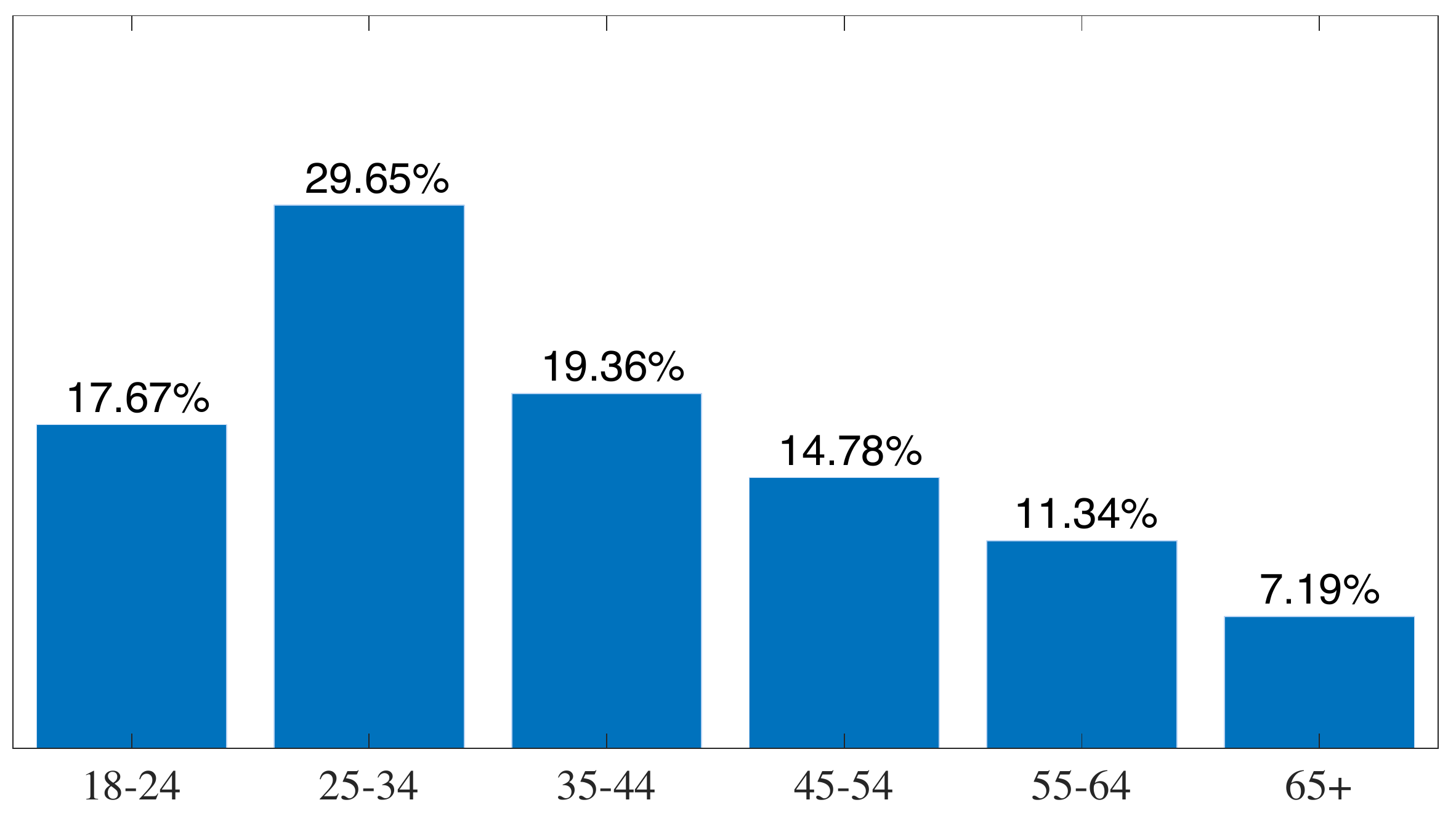}
  \caption{Age distribution on Digg.}
  \label{fig:digg_age}
\end{figure}

\begin{figure}[bth]
  \centering
  \includegraphics[width=0.9\linewidth]{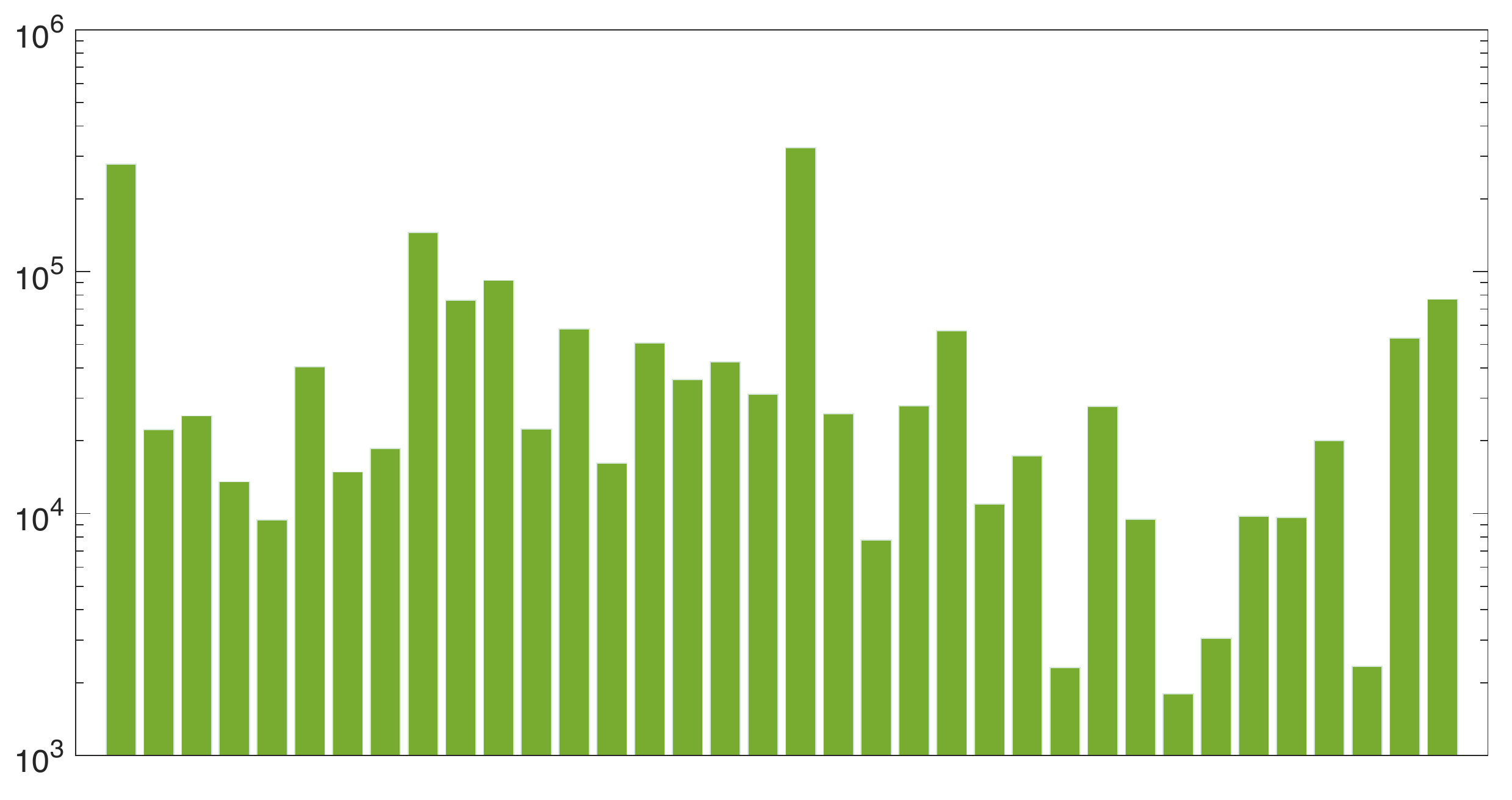}
  \caption{Location distribution on Weibo.}
  \label{fig:weibo_location}
\end{figure}

We describe here the distribution on the sensitive attributes \emph{age} in Digg and \emph{location} (\emph{region}) in Weibo, as illustrated in Fig.{\ref{fig:weibo_location}} and Fig.{\ref{fig:digg_age}}. In Weibo, there are 36 locations all together, representing the province / region of users. In Fig.{\ref{fig:weibo_location}}, each bar represent a location (names are irrelevant). The age-range distribution comes from publicly available statistics provided by Digg platform, where the users are classified into $6$ different age groups, as illustrated in Fig.{\ref{fig:digg_age}}.

\subsection{Unbalanced Weibo dataset - \emph{gender} flipping}
\label{sec:flipping}
To create unbalanced diffusion cascades, we manually modified them at the granularity of influencers, in this way simulating situations where being fair or not in information diffusion is a characteristic of the influencers, which then as a side effect leads to the sensitive attribute being unbalanced overall in the population of cascade participants.   

In Weibo, for the \emph{gender} sensitive attribute, we first randomly selected 50\% of the influencers as the working group, from which a more unbalanced \emph{gender} distribution is to be obtained.

For each selected to-be-unfair influencer, we then went over all its diffusion cascades and flipped the \emph{gender} profile for  50\% randomly selected  \emph{male} participants. We chose to modify the data in this direction, since the \emph{male} population was already slightly smaller in the initial dataset.  This had two consequences: roughly half of the influencers were now unfair for \emph{gender}, and  overall we had a \emph{male / female}  ratio, in the entire dataset, of roughly $25\%-75\%$. 
 
\begin{figure}[h]
  \centering
  \includegraphics[width=1\linewidth]{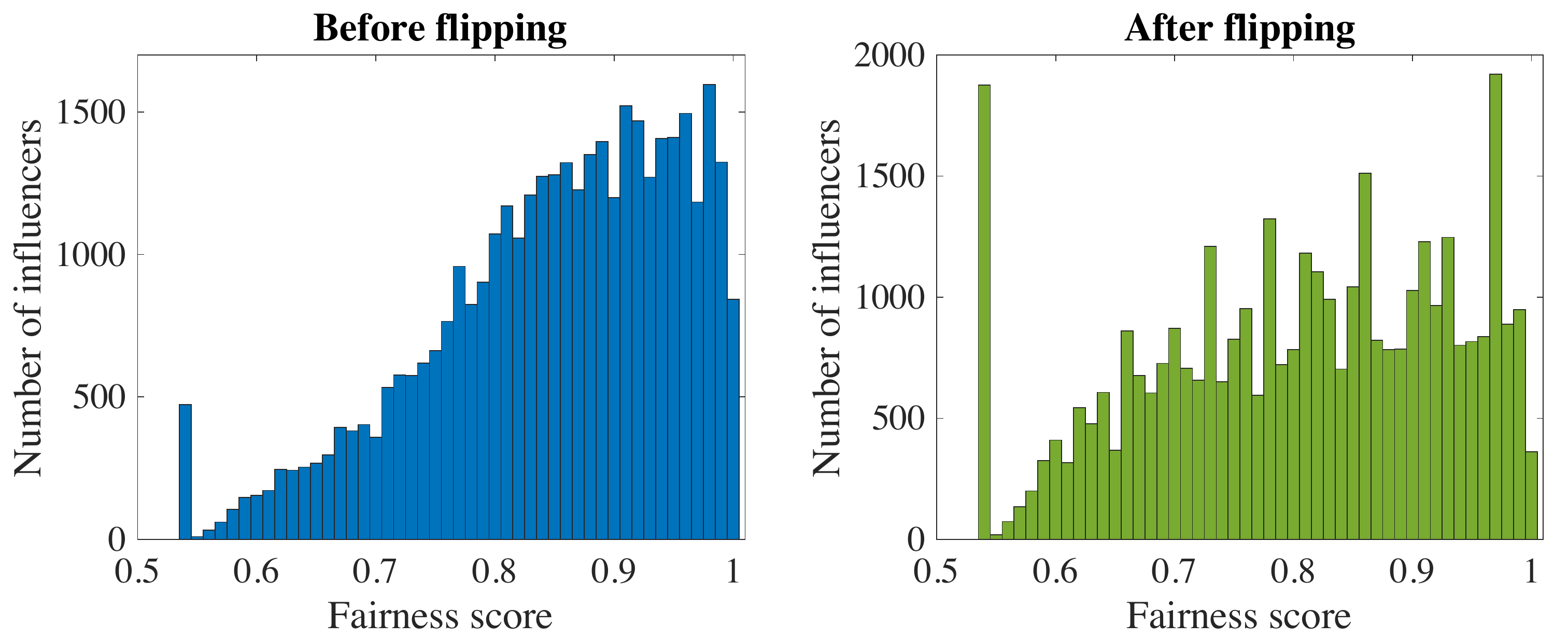}
  \caption{Fairness distribution --  \emph{gender} in Weibo.}
  \label{fig:distribution_flipping}
\end{figure}

 The impact of these changes on the per-influencer fairness score distribution is given in Fig.{\ref{fig:distribution_flipping}}, where we display the top 1000 influencers (those that matter most for our IM algorithms). We can see that, after flipping, the fairness scores decrease in general and are much more diverse.   Recall also that, by definition,  \emph{gender} fairness will always be at least $0.5$, as illustrated in the plots of  Fig.{\ref{fig:distribution_flipping}}.

\subsection{Aggregative training vs. direct combination of per-attribute node embeddings}
\begin{figure}[t!]
  \centering
  \includegraphics[width=0.87\linewidth]{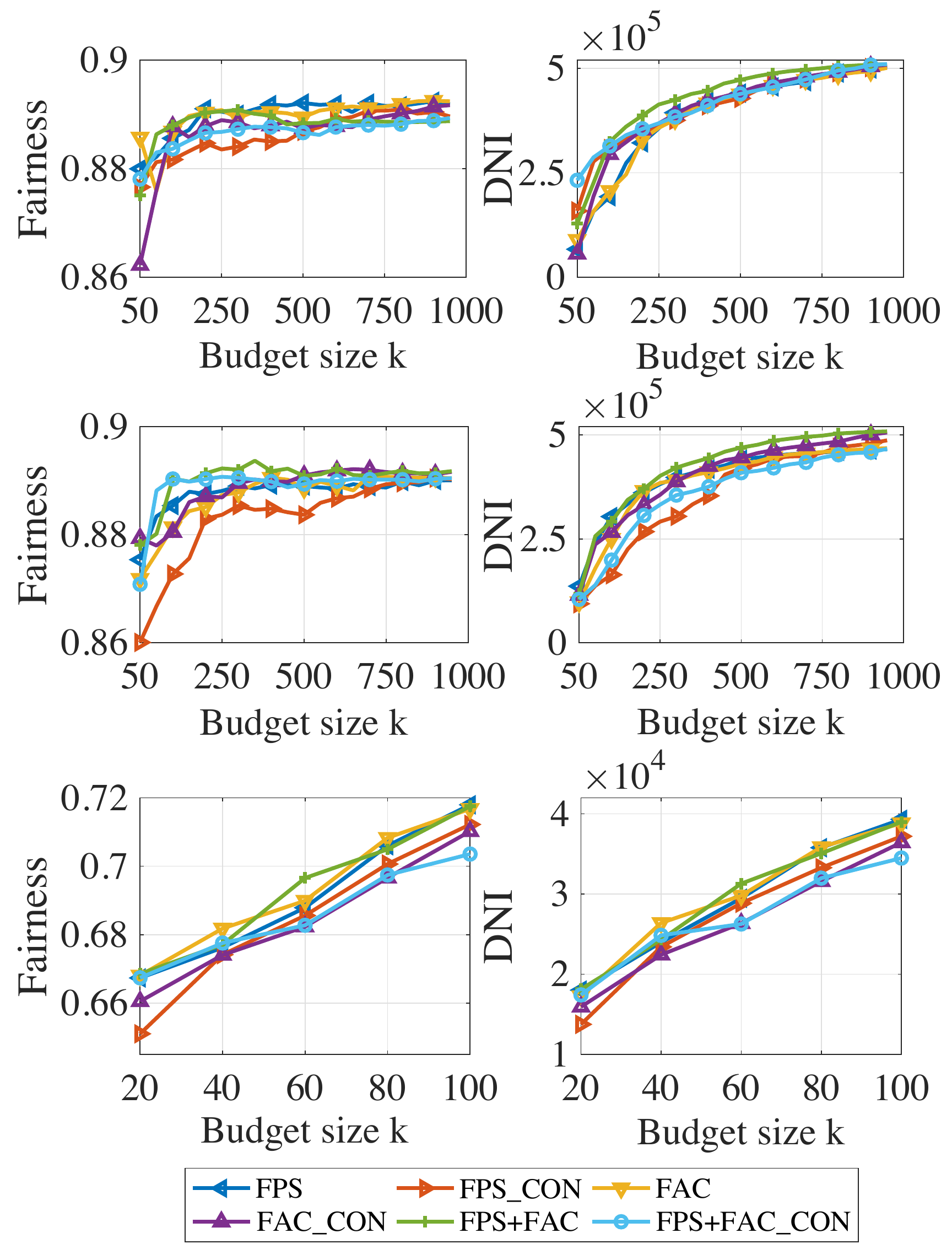}
  \vspace{-2mm}
  \caption{Aggregative training vs. concatenation of embeddings (top-2 rows: Weibo \& flipped-\emph{gender} Weibo, 3rd: Digg).}
  \label{fig:concat}
 \vspace{-1mm}
\end{figure}

\vspace{-0.5mm}
\label{aggregative-concatenation}
We consider here the following question: \emph{can we train our models on individual sensitive attributes, and then use the resulting embeddings for fair IM on combinations of attributes}? This could be very beneficial, allowing us to avoid a costly training phase whenever  many (or all) combinations of sensitive attributes may arise in fair IM queries.  We describe some initial results (Fig.\ref{fig:concat}), comparing models trained specifically on a given combination of attributes (what we call \emph{aggregative training}) with models that use directly (by concatenation) the node embeddings trained for the individual attributes of that combination.  In Digg, the embeddings obtained from by aggregative training show a clear advantage over the direct concatenation of embeddings trained for \emph{gender} and \emph{age} separately, on both fairness and influence. In Weibo, the gap between the two alternatives is reduced. One possible explanation may be that for the aggregative training in Weibo we deal with 72 categories (as opposed to 12 in Digg),  which may bring noise in the learning of node representations. As a preliminary conclusion here, the trade-off training cost vs. performance seems to be in favor of such a straightforward concatenation of the node embeddings trained for single attributes in separation. 
We leave as a future extension a thorough study on how to deal with this trade-off.  

\end{document}

%% file: introduction.tex
\label{sec:intro}

Social media plays a crucial role in connecting people in the Internet era, allowing everyone to exchange information easily. Importantly, social networks such as Facebook, Twitter, or Instagram can be used as a powerful medium to carry out information diffusion campaigns of various kinds, e.g., in marketing and political scenarios. Indeed,  social media advertising is a booming domain,  gradually replacing advertising over the more traditional channels  (TV, radio, print, mail, etc). It is enabled by the highly effective word-of-mouth mechanisms that are embedded in social applications.  

\eat{
Social networks are therefore an unprecedented medium for advertising, be it with a commercial intent or not, as news, products, political manifests, etc, can easily propagate to a large yet targeted audience.  But the interest of social media for marketers is twofold. Not only it enables to easily and rapidly reach many users, but it also brings \emph{credibility} to the messages that are being conveyed. Indeed, there are studies showing that people are more inclined to pay attention to a message or referral coming from a friend or an influencer whom he or she follows \cite{Referral1, Referral2}.  
}

Social advertising campaigns are usually carried out with a limited \emph{budget}, and so, the challenge is on how to maximize the spread of information (influence) accordingly.  The family of algorithmic problems under the generic name of \emph{influence maximization} (IM) \cite{kempe2003maximizing} pertains to all these scenarios that aim to maximize information spread in a diffusion network under budget constraints. It aims to select the $k$ most influential nodes from which the diffusion of a specific message should start.  In many ways, IM  mirrors an increasingly popular and  effective form of marketing in social media, which targets a sub-population of \emph{influential people}, instead of the entire base of users of interest, known as \emph{influencer marketing} \cite{InfluencerMarketing}.

Furthermore, recent trends have motivated advertisers to seek more control over who may receive their messages, as opposed to simply maximizing the spread of information. Along this line, there is an increasing need for information diffusion in a fair way. Indeed, many campaigns -- e.g.,  sharing public opinions, job or loan advertisements, news sharing, etc --  may require not only a large spread, but also \emph{fairness} in influencing users across various sensitive attributes like race, gender, or location, in order to achieve unbiased viewpoints. Therefore, a recent focus has been on how to maximize fairness along with the influence spread \cite{DBLP:conf/aaai/RahmattalabiJLV21,DBLP:conf/www/FarnadiBG20, DBLP:conf/aaai/BeckerDGG21,gaps2019,groupinfluencemax2019}. Yet the challenge remains on how to achieve a scalable solution for fair influence maximization, while enabling a flexible formulation that can support one or multiple sensitive attributes.

\eat{
Thus, social networks like Facebook, Twitter, Wiebo, etc. are powerful media to influence public opinion for various economical, political, and welfare events. Influence of a source user on target users depends on the interconnectivities between the users and their involvement into various activities, such as originating or participating in information spread. The possibility of a target user adopting behavior from a source user through social media is formally defined as social influence and the expected number influenced users is formally measured as influence spread. Finding a set of source nodes, also known as influencers, maximizing the influence spread is a well-known influence maximization problem \cite{maxspreadinfluence2003}. The influence maximization problem has been getting a lot of attention lately with the increasing applications of social media to successfully carry out various campaigns like viral marketing, epidemic spread, sharing publish opinion, social recommendations etc.. Usually, such campaigns are executed with limited resources and budget, and thus, the number of influencers to be selected in the influence maximization problem is known a priori. Furthermore, campaigns like disseminating news, collecting public opinion, and broadcasting job requisites aim to find a set of target nodes, also known as infectors, is diverse with respect to sensitive attributes, such as age, gender, race, location, etc.. Thus, the challenge in the influence maximization project is how to maximize fairness along with the influence spread \cite{onthefairness2019}.
}

In this paper, we generalize the influence maximization problem by modeling the objective as a function of spread and fairness across a set of sensitive attributes. To quantify fairness, we propose an analytical model supporting a set of attributes with n-ary categories. For this problem, by exploiting the historical diffusion cascades and the users’ fairness  information, we then propose two deep-learning based solutions: (a) fairness-based participant sampling (FPS), and (b) fairness as context (FAC). Importantly, we achieve scalability by building our predictive models from the observed diffusion cascades, instead of relying on computationally intensive simulations of diffusions. In \replaced[id=y]{FPS}{the former approach}, node embeddings and diffusion probabilities are derived while considering fairness and information cascade independently. On the other hand, \replaced[id=y]{FAC}{the latter approach} adopts \replaced[id=y]{joint learning}{ correlations} between fairness and information cascades, in order to derive the node embeddings and diffusion probabilities. Finally, in both algorithms, diffusion seeds are \deleted{greedily} selected \added[id=y]{through a fair-greedy algorithm} using the node embeddings and diffusion probability information.

In summary, our main contributions are as follows: 
\begin{itemize}
    \item 
    We propose a new formulation for fair influence maximization  and a model supporting a set of categorical attributes to quantify fairness.  Our model is based on the notion of equity, i.e., ensuring that nodes from different groups of interest are (almost) equally likely to be influenced.
    \item
    We describe two algorithms for this problem, called FPS and FAC. These deep-learning based solutions are applicable to arbitrary sets of sensitive attributes and can scale \added[id=y]{to large, realistic social graphs}. We achieve scalability by exploiting the diffusion cascades information to solve the fair influence maximization problem, as opposed to social connectivity and computationally expensive diffusion simulations.
    \item
    \added[id=y]{We propose a model-independent fairness-aware greedy algorithm to select the seed set that maximizes influence, based on the learnt graph representation, which comes with a knob to control the degree of aversion to inequality.}
\end{itemize}

 We evaluate these algorithms over a real-world and publicly available  dataset (Weibo), \added[id=y]{and a hybrid real-synthetic dataset (Digg), in which we randomly allocate user features  based on real feature distributions.} \added[id=y]{By comparing with the state-of-the-art methods, we show that FPS and FAC}
 \deleted{showing that they} can maximize the number of activated nodes while achieving the required fairness w.r.t. the sensitive attribute(s). \deleted{By comparing FPS and FAC with baseline methods based on state-of-the-art ideas, we show that they outperform these methods in terms of spread, fairness, and scalability.}

 \eat{
The paper is organized as follows. We first discuss the main related work. The fair influence maximization problem is then formally defined and the proposed algorithmic solutions are described in the subsequent section. The performance of our algorithms is then empirically evaluated, prior to concluding the paper. 
}

%% file: related.tex
\label{sec:related}

\textbf{Influence maximization} (IM) usually has as objective the \emph{expected spread} under a stochastic diffusion model, which describes diffusions as probabilistic processes.  The work of \cite{kempe2003maximizing} introduced two  such discrete-time diffusion models, Linear Threshold and Independent Cascade, which  have been adopted by most of the literature that followed (see for instance the survey of \cite{8295265}). Such models rely on diffusion graphs with edges weighted by a score of influence, i.e., a spread probability. Since selecting the seed nodes maximizing the expected spread is NP-hard under such diffusion models, greedy / approximation graph algorithms that rely on the monotonicity and sub-modularity of the objective have been studied extensively. However, scaling IM algorithms to large, realistic networks remains difficult. Indeed, most of the IM research focuses on improving  efficiency and scalability (see  the benchmarks of  \cite{DBLP:conf/edbt/0001GR19,DBLP:conf/sigmod/AroraGR17}).  

\vspace{1mm}
\textbf{Influence maximization with cascades.} IM studies may also differ in the underlying data assumptions. The work of \cite{socialinfluencemax2018} has argued recently that algorithms based on stochastic models that rely only on the connectivity information lead to sub-optimal results, as information related to the users’ diffusion activities (information cascades) is ignored.  In other studies (e.g., \cite{scalableinfluence2013,Saito2009LearningCI}), the authors considered exploiting the historical cascades along with social connectivity information, in order to derive probabilities of influence under IC or LT assumptions. However, the simplifying assumption of diffusion independence  may lead to inaccurate estimation since the correlations across participating nodes in the cascades are ignored. Recently, \cite{representationlearning2016, multitask_2020} addressed  IM by extracting influencer and susceptible \emph{embeddings} from historical cascades and used such learned models to make spread predictions and  to solve IM. Such deep-learning based solutions avoid the independence assumption and consider high-order correlations between users.

\vspace{1mm}
\textbf{Graph algorithms for fair IM.}  As the gap between theoretical models /  algorithms for IM and real-world diffusion scenarios with realistic assumptions reduces gradually, recent studies have considered the \emph{fairness} of diffusion campaigns, by revisiting graph algorithms for IM. In \cite{groupinfluencemax2019}, the authors describe two initial notions for group fairness, \emph{maximin} (inspired by the legal notion of disparate impact) and \emph{diversity} (every group receiving influence commensurate to what it could have generated on its own). They show that these fairness objectives  -- while non-submodular --  can be reduced to multi-objective sub-modular optimization, and the authors describe a method for general multi-objective sub-modular optimization.  In a similar vein,  \cite{gaps2019} studies  the maximin criterion,  presents hardness results, and ana\-lyses several adaptations of the IM greedy strategy. Similar to the two aforementioned works, \cite{DBLP:conf/aaai/BeckerDGG21} focuses on maximin fairness,  considering probabilistic strategies for seed selection, which leads to approximating solutions for fair IM with provable guarantees, close to those for classic IM. 

For a broader scope, in \cite{DBLP:conf/www/FarnadiBG20}, a generic framework based on integer programming is shown to capture various flavors of group fairness in IM such as \emph{equality}, \emph{equity}, \emph{maximin}, or \emph{diversity}. As these cannot be jointly enforced, in practice one must chose based on the application scenario. \cite{DBLP:conf/aaai/RahmattalabiJLV21} also provides a broad formal study on desirable fairness properties by (i) transposing existing principles (e.g., monotonicity) from social welfare theory on the ``utility vectors'' that describe spread across groups of interest, and (ii) proposing new principles specific to group fairness, such as the \emph{utility gap reduction}. They propose a model of welfare functions, which allows to tune the fairness vs. spread  trade-off and enables a generic algorithm, applicable for various fairness definitions. Impossibility results are also discussed,  such as for jointly obeying the utility gap reduction along with some of the other principles.

Among other related graph-based algorithms on fairness and information diffusion,  \cite{onthefairness2019} considers  a time-constrained spread formulation, along with IM algorithms for an objective function that combines spread and maximum disparity. In \cite{DBLP:conf/aaai/BeckerDGG21}, the focus is on  maximizing the diversity of the information that reaches users, instead of the diversity of the users who get a particular piece of information. As such, this is orthogonal to the goals of fair IM. \cite{DBLP:conf/www/StoicaHC20} studies how the diversity of the seed set improves fairness of spread.  \cite{DBLP:conf/wsdm/LinLL20} takes another approach for fair IM, by sampling first under fairness constraints from the initial graph.  The work of \cite{DBLP:conf/edbt/GershteinMY21} starts from the observation that maximizing the influence for one group may come at the cost of reducing it for other groups. They propose a model allowing to explicitly choose the desired balance between the groups' spread objectives, where all objectives except one are turned into constraints, while the remaining objective is optimized subject to the constraints. In short, constraints stipulate that any IM solution must yield at least a given fraction of the optimal influence.

 These graph algorithms for fair IM have major limitations, that we aim to address in this paper, such as (i) working exlusively with the diffusion topology, often completed by synthetic diffusion probabilities, which has limited practical relevance, 
 (ii) often relying on fairness criteria that in practice may still lead to unbalanced spread,  and (iii) failing to achieve a good trade-off between fairness and spread in a scalable manner. Indeed, scalability remains a key issue for practical purposes, as discussed for instance in \cite{DBLP:conf/www/FarnadiBG20}.  The reason is that these state-of-the-art  approaches work by (i) adapting existing IM algorithms for greedily selecting seeds from the diffusion graph, and (ii) by relying on computationally expensive  procedures (e.g., Monte-Carlo or Reverse-Influence sampling). Moreover, such adaptations must also cope with the fact that fairness-aware IM no longer benefits from the submodularity of the influence objective.

\vspace{1mm}
\textbf{Learning embeddings for fair IM.} By exploiting cascades and node features -- two data facets often available in  diffusion scenarios -- we propose  to \emph{learn} node embedding models allowing us to solve instances of the fair IM problem in a manner that is both efficient and flexible w.r.t. the spread effectiveness vs. fairness trade-off. 

  Along this research path -- learning node representations for fair IM -- the works that are closest to ours are \cite{adversialgraph2020,khajehnejad2022crosswalk}. In \cite{adversialgraph2020}, the authors propose an adversarial neural network-based approach. First, a set of influencers are identified using a variant of K-means clustering over node embeddings obtained by the adversarial neural network. Then, the IC model is used to identify a diverse  influencer set. 
 The follow-up work \cite{khajehnejad2022crosswalk} revisits generic algorithms for learning node representations (e.g., DeepWalk \cite{perozzi2014deepwalk}, node2vec \cite{grover2016node2vec}) in order to ``boost'' fairness. This is done by biasing random-walks in the vicinity or across  boundaries between the  user groups for which fairness must be achieved.  By assigning larger weights to edges that are at groups' peripheries, 
  they obtain embeddings by which  the selected seeds can spread the information more effectively to multiple groups. The limitations of \cite{adversialgraph2020,khajehnejad2022crosswalk} are threefold: (i) as they exploit the social connectivity, they incur a high computation cost and cannot reasonably scale; in particular, training the adversarial neural network in \cite{adversialgraph2020} grows exponentially with the number of users in the social network, (ii) they assume that the probability of diffusion between a pair of nodes is constant and follows the Markovian assumption, and (iii) they are not generic, i.e., not applicable for fairness across any and multiple sensitive attributes.

Compared to \cite{adversialgraph2020, khajehnejad2022crosswalk},  we revisit IM with fairness under a more flexible formulation, 
by algorithmic solutions that 
are applicable to  arbitrary sets of sensitive attributes and  to large, realistic datasets. Our methods alleviate therefore the main practical limitations of the state-of-the-art solutions, including the one of scalability.  

Finally, fairness-aware deep representation learning -- e.g., for users in a recommender system \cite{wu2021fairness} -- have been proposed for unbiased representations of users, s.t. sensitive attributes cannot be inferred from the model's output.  Here, the neural networks are used to learn
 the influencing and fairness aptitude of influencers from the cascades, aspects which pertain to the user base (population) that may be influenced, and not to the influencers themselves.

\deleted{We evaluate the performance of the proposed algorithms, by comparison with both the IMInfector algorithm of \cite{multitask_2020} -- the deep-learning approach focusing on spread maximization -- and the adversarial graph-based Fair-Embedding algorithm of \cite{adversialgraph2020},  over the public Sina Weibo social network dataset. We stress that the interest of this dataset, designed specifically for information diffusion studies, is that it can lead to verifiable and reproducible benchmarking indicators.  The empirical results show that our solutions outperform these state-of-the-art methods,  achieving the right (chosen) balance between the proportion of influenced nodes and the fairness across the network.}

%% file: preliminaries.tex
\label{sec:preliminaries}

In the fair IM literature, akin to algorithmic fairness studies, several notions of \emph{group} fairness have been studied. We briefly discuss them next, motivating our fairness model choice in the process.

\begin{itemize}
    \item \emph{Equality}: conceptually straightforward, it refers to the fair allocation of seeds by the IM algorithm, i.e., each group getting the same number of seeds. Therefore, this design does not necessarily prevent unfair spread.
    \item \emph{Maximin}: its goal is to minimize the gap between groups / ca\-tegories w.r.t. the received influence relative to their size. Focusing on the  minimum influence among the groups relative to size may drastically diminish the overall spread, e.g., when one group is less well-connected than the others. 
    \item \emph{Diversity}: the crux of this fairness model is to ensure that each group / category of users receives influence at least equal to their internal spread, i.e., in the graph induced by that group. Under this formulation, diversity is not ``inequality averse'', as it may lead to unbalanced spread.  
    \item \emph{Equity}: similar to maximin, equity is a model that aims to reach a fair share of influence, this time by ensuring that any node's probability of getting influenced is (almost) the same, regardless of the group it belongs to. So a group's expected number of influenced nodes should be proportional to its ratio in the graph. Equity corresponds to what is also known as \emph{demographic parity} (DP) in  algorithmic fairness studies.  
\end{itemize}
Our fairness model is based on the equity notion, which strictly corresponds to the ideal fairness for a sensitive categorical attribute $s$ (or a combination thereof) and the groups it induces in the graph by $C_s$, \added[id=y]{the set of categories of $s$.} \deleted{(Eq. (\ref{eq:4})). Then, for generality, across different sensitive attributes that may be of interest, we assume a weighted combination for the overall fairness $g$ (Eq. (\ref{eq:2})). Furthermore, equity / exact demographic parity is the only principle that obeys the utility gap reduction
principle \cite{DBLP:conf/aaai/RahmattalabiJLV21}, which states that, among different spread outcomes, we should prefer the one whose total utility is highest and also has the smallest utility gap.}
\deleted{Next, we describe in detail our proposed algorithmic solutions to maximize jointly influence spread and fairness.}
\added[id=y]{To evaluate the fairness of a set of selected influencers with respect to a set of sensitive attributes, we next derive an analytical expression to quantify fairness.}

\vspace{1mm}
\textbf{Proposed fairness model.}
\added[id=y]{Under the notion of equity,} we consider categorical attributes that are of interest in achieving fairness. \deleted{For each sensitive attribute $s \in S$ (e.g., \emph{gender}) we may have multiple categories, which we denote by $C_s$. In short,} In what follows, we discuss fairness for a single sensitive attribute $s$, but this extends easily to a set (combination) of categorical attributes. For a fair  spread, we want the proportion of users belonging to a particular category (or \emph{group}) of $s$ in the entire population to remain the same after that spread, i.e., in the influenced population.  For example, if we have a $70\%$ \emph{male} population (the groups being \emph{male} / \emph{female}), we want, after the spread, the influenced nodes to be made of $70\%$ \emph{males}. This would be the ideal fairness per sensitive attribute and categories thereof. However, if we cannot achieve this, we need to assess the deviation from the ideal situation.  In a spread context, for each categorical value $i \in C_s$ (e.g.,  \emph{\replaced{gender: female}{origin: European}}) we will have two sets of users of interest: (1) the users $V_i \subseteq V$ belonging to category $i$ in the entire population (e.g., \emph{all \replaced{women}{Europeans} in } $V$), and (2) the influenced users (after the spread) belonging to category $i$, denoted by $\Omega_i \subseteq V_i$  (e.g.,  \emph{all influenced \replaced{women}{Europeans}}). 

For an attribute $s$ with multiple categories $C_s$, the ideal fairness of the diffusion results would be
\begin{equation}\label{eq:perfect_fair}
    \frac{\left|\Omega_i\right|}{\left|V_i\right|} \approx \frac{\left|\Omega_j\right|}{\left|V_j\right|} \approx \frac{\left|\Omega_k\right|}{\left|V_k\right|} \approx \dots \forall i, j, k, \dots \in C_s.
\end{equation}
We have perfect fairness when the proportion of influenced nodes in all the categories (groups) of $s$ is the same. Hence the \emph{coefficient of variation} $CV$ can be used here to evaluate the degree of variation (unfairness) of influenced ratios for all the categories,  computed as:
\begin{equation}
    CV = \sigma / \mu,
\end{equation}
where $\sigma$, the standard deviation of the influenced ratios, 
is: 
\begin{equation}
    \sigma = \sqrt{\frac{\sum_{i \in C_s}\left(\frac{|\Omega_i|}{|V_i|}-\mu \right)^2}{|C_s|}},
\end{equation}
and $\mu$ denotes the average of influenced ratios: 
\begin{equation}
    \mu = \frac{1}{|C_s|}\sum_{i \in C_s}\frac{|\Omega_i|}{|V_i|}.
\end{equation}
With the relative dispersion of influenced users in the groups induced by $s$ to capture unfairness, the fairness score can then be  scaled by a \emph{sigmoid} function and bounded between $0$ and $1$, by\footnote{By definition, fairness is above $\sqrt{n-1}/n$, for $n$ the number of categories of the sensitive attribute or combination thereof. E.g., \emph{gender} fairness is always above $0.5$.}
\begin{equation}
\label{eq:fairness}
       f_s = \frac{2}{1+\exp(CV)}.
\end{equation}

\deleted{We define fairness $f_s$ for one sensitive attribute $s$ in Eq, as a quantity between $0$ to $1$. In order to take into account all the categories of $s$, 
in the numerator of $f_s$, within each category $C_s$, we are looking at the ratio between influenced users and users in the entire population. As the denominator, we have a normalization factor. We want fairness w.r.t. the sensitive attribute $s$ to reach the maximum (value $1$) when all the categories of $s$ have exactly the same ratio between influenced users and the entire population.}

%% file: problemdef.tex
\label{sec:probdef}
We formulate the fair IM problem to optimize both influence spread and fairness as follows. We are given a social network $G(V, A)$, where $V$ is a set of nodes representing social network users and $A$ is the set of edges representing the social connectivity between users (e.g., derived from users’ contacts and following information). Additionally, we have a set $S$ of categorical, sensitive user attributes, such as \emph{gender}, \emph{race}, \emph{location}, etc., which are of interest when aiming for spread fairness. Each node (user) will have a combination of categories  pertaining to this set of attributes associated to it.  
\deleted{Assuming that not all sensitive attributes may be equally important for spread fairness, we also associate an importance weight $w_s$ to each attribute  $s$. }

Besides the social structure and the users' sensitive attributes, we also represent the information from \emph{diffusion cascades} $D$ (logged traces of information spread in the network),  with each cascade $d\in D$ being a set of pairs $(v, t_{v})$ where $t_{v}$ is the timestamp when the information was diffused by user $v \in V$. In each cascade $d\in D$, the user having the smallest timestamp is the initiator of the cascade, while all the other users are participants. Let also $I$ denote the overall set of initiator nodes (influencers) from the cascades $D$.

\added[id=y]{As  diffusions do not necessarily follow the so\-cial topology, we aim to learn and represent diffusion probabilities between nodes based on the cascades, with both spread and fairness as objectives.}  Our problem is to find a set of $k$ influencers that maximizes jointly the spread (number of influenced nodes) and its fairness. 
\deleted{and the weighted fairness for sensitive attributes across the influenced nodes.
For a given importance factor $0 \leq \alpha \leq 1$, defining how much valuable influence spread is w.r.t. weighted fairness across the sensitive attributes, the objective  is to solve the following optimization problem:}

\eat 
{\begin{equation}\label{eq:1}
\argmax_{I' \subseteq I, |I'|\leq k}
      \left\{\alpha \times \left( \frac{m}{\hat{m}}\right) + (1-\alpha) \times \left( \frac{g}{\hat{g}}\right)\right\}
\end{equation}
\begin{equation}\label{eq:2}
    g = \frac{w_{s}f_{s}}{\sum_{s \in S} w_{s}}
\end{equation}
}

\deleted{where $m$ denotes the number of influenced nodes in the outcome of a spread initiated at the seed set $I'$ and $\hat{m}$ is an upper bound on the number of influenced nodes. Similarly, $g$ denotes weighted fairness across sensitive attributes in the spread outcome as defined in Eq. (\ref{eq:2}),  and $\hat{g}$ denotes an upper bound on the weighted fairness. }

\deleted{Next we define $f_{s}$, the fairness of a sensitive attribute $s$, and we describe the analytical model to quantify it. In our objective function from Eq. (\ref{eq:1}), the influenced node count and the fairness are normalized by their upper bounds, and thus, the objective value ranges between $0$ and $1$. Both upper bounds are data-based (computed from the underlying data). For space reasons, the details on how we compute these upper bounds on influence spread $\hat{m}$ and fairness $\hat{g}$ can be found in the appendix (supplementary material).}

%% file: solution.tex
\label{sec:solution}


\begin{figure*}[t!]
  \centering
  \includegraphics
  [width=0.85\linewidth]
  {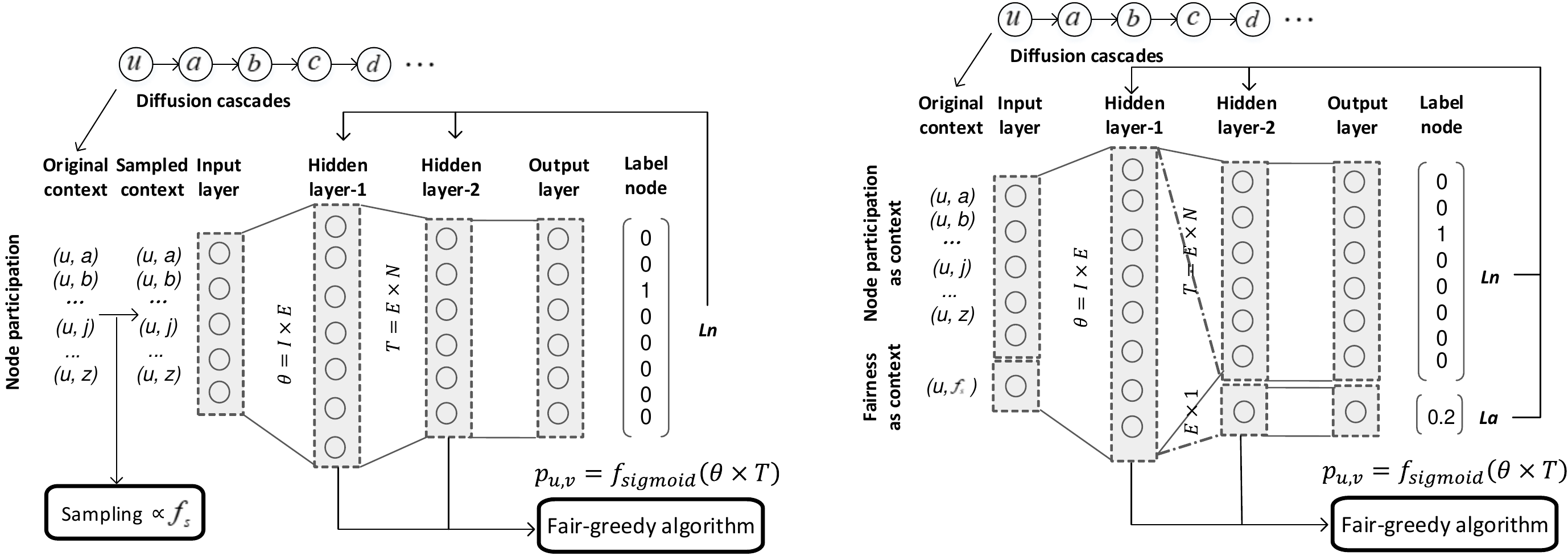}
  \vspace{-2mm}
  \caption{Illustration of our algorithmic solutions: FPS (left) and FAC (right).}
  \label{fig:fps_fac_framework}
\vspace{-2.5mm}
\end{figure*}

We propose two algorithms, Fairness-based Participant Sampling (FPS) and Fairness As Context (FAC)\footnote{Code and experiments  available at \url{github.com/goldenretriever-5423/fair_at_scale}.
}, to address the fair IM problem. The crux of these proposals is to extract the influencing aptitude and fairness of users from  cascades and to represent this information in a multidimensional space using a neural network. The representation of users' features in a latent space is referred to as node embedding \cite{inf2vec}. Moreover, diffusion probabilities between  pairs of cascade initiators (influencers) and cascade participants (influencees) are derived using the neural network. Finally, the node embedding and diffusion probabilities are used to select a set of influencers by a \added[id=y]{fair-}greedy approach. In the FPS approach, node embedding and diffusion probabilities are derived by considering users' spread  aptitude and fairness  independently. FAC jointly considers users' spread aptitude and fairness,  to derive node representations and diffusion probabilities. We describe these algorithms next,  w.l.o.g   limiting the description to a single sensitive attribute $s$. 

\vspace{-0.5mm}
\subsection{Fairness-based Participant Sampling (FPS)}
\vspace{-0.5mm}
Information about the influencers' aptitude in influencing others can be derived from historical cascades. FPS explicitly penalizes the spread aptitude of an influencer as a function of the influencer's fairness. Fig. \ref{fig:fps_fac_framework} (left) shows the schematic representation of the FPS model. The original context is exclusively extracted from historical cascades, where the input feature is the initiator of a cascade and the output labels are the set of participating nodes in the cascades. 
 The key idea is to first oversample the original context of a cascade $d\in D$ by $\eta\%$ and then downsample the oversampled context based on (a) the temporal information of actions taken by the participating nodes in the cascade, and (b) fairness of the cascade initiator. The rationale is  that quickly responsive participating users are highly likely to be influenced. Thus, the probability of sampling a node $v$ is inversely proportional to the elapsed time $t_v - t_u$ since $d$ was initiated by node $u$. From the cascade $d$, the probability of sampling node $v$ from the initiator node $u$ is defined as follows:
\begin{align}\label{eq:5}
    p\left( v|d\right) = \frac{\left(t_v - t_u\right)^{-1}}{\sum_{(x,t_x) \in d}\left(t_x - t_u\right)^{-1}}
\end{align}
Let $f_{u}^{s}$ be the fairness of an influencer $u$ \added[id=y]{w.r.t. a sensitive attribute $s$}, obtained as in Eq. (\ref{eq:fairness}) when restricting to the cascades initiated by $u$. The context is further downsampled by selecting $|d| \times f_u^s$ participating nodes, where $|d|$ is the length of cascade $d$ \added[id=y]{and the fairness score is used as a penalty factor to downsample the influencers who started unfair information diffusions w.r.t. the attribute $s$.}

Next, the sampled context is input to the deep neural network, to learn about the influencers' spread aptitude and the diffusion pro\-ba\-bilities for influencer-influencee pairs. The input context for a given  cascade $d$ with sample length $l$ is $X_d=\left\{\left(x, y^1\right), \dots, \left(x, y^l\right)\right\}$, for $x \in R^{|I|}$ and  $y^j \in R^{|V|}$. Recall that  $I$ is the set of cascade initiators in the training dataset and $V$ is the set of all the nodes in the social graph. A node in the context is represented using one-hot encoding. The first hidden layer contains the number of neurons equivalent to the embedding size $|E|$, and thus, the parameters at the first hidden layer are $\theta \in R^{|I| \times |E|}$. The number of neurons at the second hidden layer is equivalent to the number of nodes in the social network, and thus, the parameters in the second hidden layer are $T\in R^{|E| \times |V|}$. The number of nodes at the output layer is $|V|$ with the softmax activation function.   

The network is trained to minimize the cross-entropy loss\footnote{We use Noise-Contrastive Estimation (NCE) loss, which is efficient when dealing with such large softmax
problems; it is fundamentally a negative sampling method to approximate the multi-class problem to a binary one.}  function $L^n$. Eq. (\ref{eq:6}), (\ref{eq:7}), (\ref{eq:8}) give the analytical operations of the model. At iteration $n$,  $z_u^n$ denotes coefficients at the output of second hidden layer for influencer $u$, and $o^n_u$ is the softmax transformation at the output layer for user $u$. 
\begin{equation}\label{eq:6}
    z_u^n = \theta_u \times T + b^{n}
\end{equation}
\begin{equation}\label{eq:7}
    o_u^n = \frac{e^{z_u^n}}{\sum_{v \in V} e^{z_v^n}}
\end{equation}
\begin{equation}\label{eq:8}
    L^n= Y^n\log(O^n)
\end{equation}
\eat{
\pv{Fix the formatting}
\begin{table}[ht]
\caption{Models of the proposed neural networks}
\centering
\begin{tabular}{|| c | c ||}
\hline
FPS approach & FAC approach \\
\hline
\hline
\begin{equation}\label{eq:6}
    z_u^n = \theta_u \times T + b_{n}
\end{equation}
&
\begin{equation}\label{eq:13}
    z_u^n = \theta_u \times U + b_{n}
\end{equation}
\\
\hline
\begin{equation}\label{eq:7}
    o_n^u = \frac{e^{z_u^n}}{\sum_{v \in G} e^{z_v^n}}
\end{equation}
&
\begin{equation}\label{eq:14}
    o_n^u = \frac{1}{1+ e^{-z_u^n}}
\end{equation}
\\
\hline
\begin{equation}\label{eq:8}
    L_n = Y_n\log(O_n)
\end{equation}
&
\begin{equation}\label{eq:15}
    L_n = \left( G_n - O_n \right)^2
\end{equation}
\hline
\end{tabular}
\end{table}
}
The parameters of the first hidden layer represent the influence embedding of cascade initiators. The second norm of an embedding captures the aptitude of a user in influencing other users. Applying the fairness-based penalty in generating contexts penalizes the spread aptitude of biased influencers. On the other hand, the parameters of the second hidden layer represent the susceptible embedding of users. Applying the sigmoid transformation to the  product of both, influence and susceptible embedding results in diffusion probabilities $p_{u,v}$ that are defined as follows:
\begin{align}\label{eq:9}
    p_{u, v} = f_{sigmoid}(\theta \times T).
\end{align}
A diffusion probability $p_{u,v}$ is therefore the probability of node $v$ appearing in cascades started by $u$. Such probabilities can be used to identify influenced nodes, with the assumption that there exists direct or indirect connectivity between users $u$ and $v$. Theferore, unlike exploring all possible connections between users $u$ and $v$ \cite{inf2vec}, the implicit assumption of path existence drastically reduces the computational complexity, and  the algorithm becomes scalable.


\vspace{-1mm}
\subsection{Fairness as Context (FAC)}
\vspace{-0.5mm}
The FAC approach jointly considers the fairness of influencers and the size of  cascades initiated by them to derive embeddings and diffusion probabilities. As shown in Fig. \ref{fig:fps_fac_framework} (right), the model contains two separate neural networks with a common hidden layer, where the common hidden layer captures the users' embeddings. The goal of the top neural network is to capture the influence aptitude of users, in the form of embedding and diffusion probabilities. The input to the neural network is one-hot encoded sampled contexts. Unlike the FPS approach, the original context is just sampled based on the temporal information of actions taken by users, using Eq. (\ref{eq:5}). For a given cascade $d$ with length $l$, the input context is $X_d=\left\{\left(x, y^1\right), \dots,  \left(x, y^l\right)\right\}$, for $x \in R^{|I|}$ and $y^j \in R^{|V|}$. The first hidden layer contains the number of neurons equivalent to the embedding size $|E|$, and thus, the parameters at the first hidden layer are $\theta \in R^{|I| \times |E|}$. The number of neurons at the second hidden layer is equivalent to the number of nodes in the social network, and thus, the parameters in the second hidden layer are $T\in R^{|E| \times |V|}$. The number of \replaced[id=y]{neurons}{nodes} at the output layer is $|V|$ with the softmax activation function. The first neural network is trained with the objective of minimizing the cross entropy function $L^n$; its analytical operations are described in Eq. (\ref{eq:6}), (\ref{eq:7}), and (\ref{eq:8}).

The bottom neural network aims to capture the effects of fairness in the form of embedding and diffusion probabilities. The input to the neural network is $Z_j=\left(x_j, f_j^s\right)$, $x_j \in R^{|I|}$ and $f_j^s \in R^{|I|}$, where $f_j^s$ is the \deleted{weighted average} fairness of influencer $j$ w.r.t the attribute $s$, as defined before. 
 The first hidden layer contains the number of neurons equivalent to the embedding size $|E|$, and thus, the parameters at the first hidden layer are $\theta \in R^{|I| \times |E|}$. The number of neurons at the second hidden layer is one, and thus, the parameters in the second hidden layer are $T\in R^{|E| \times 1}$. The output of the neural network contains a single neuron with sigmoid activation function. The objective of the second neural network is to minimize the mean square error, $L_a$. The analytical operations of the bottom neural network can be described as follows: at iteration $n$, $z_u^n$ denotes coefficients at the output of second hidden layer for influencer $u$, $o^n_u$ is the transformation at the output layer for influencer $u$, $F^n$ is the fairness of influencers in a vector form, and $L_a^n$ is the loss function.
\begin{equation}\label{eq:13}
    z_u^n = \theta_u \times U + b^{n}
\end{equation}
\begin{equation}\label{eq:14}
    o^n_u = \frac{1}{1+ e^{-z_u^n}}
\end{equation}
\begin{equation}\label{eq:15}
    L_a^n = \left( F^n - O^n \right)^2
\end{equation}
Recall that, in the FAC architecture, the first hidden layer is common across both neural networks, where its parameters represent the users' embeddings. The key idea for designing a common hidden layer is to generate embeddings, while capturing the correlations between users' spread aptitude and fairness related features. \deleted{To avoid the overshadowing effects from the top neural network, the parameters of the bottom neutral network are updated by $l \times g_u^\beta$, where $\beta$ is a hyper-parameter and $l$ represents the cascade length. }
After training the neural networks, the embeddings and diffusion probabilities are used by the greedy influencer selection algorithm 
discussed before, to identify the set of influencers.

\paragraph{Complexity}
As each cascade is handled independently, the complexity for the sampling in FPS and FAC is $c*l$, so a linear complexity. The complexity of the neural network training is approximately $O(c*l*|V|)$, where $c$ is the number of cascades, and $l$ is the average cascade length.



\vspace{-0.5mm}
\subsection{Fair-greedy influencer selection}
\label{sec:greedy_algo}
\vspace{-0.5mm}
By the FPS and FAC models, the norm of the learned embeddings captures the aptitude of an influencer to spread information to many nodes and in a fair way, while the inner product of the embeddings for an influncer-influencee pair gives the diffusion probability thereof.  Based on these two ingredients, we describe in Algorithm {\ref{alg:fair_greedy}} our fairness-aware  greedy algorithm,  which selects the $k$ influencers who jointly maximize influence and fairness for the chosen sensitive attribute(s). Alg. {\ref{alg:fair_greedy}} adapts \cite{multitask_2020}'s  IMINFECTOR approach for influence maximization based on node embeddings learned from cascades (itself an adaptation of the CELF algorithm for IM \cite{DBLP:conf/kdd/LeskovecKGFVG07}), by adding the fairness dimension to it.    

In Alg. {\ref{alg:fair_greedy}}, $\bm{D}\in \mathbb{R}^{|I| \times |V|}$ is the diffusion probability matrix, $\bm{F}\in \mathbb{R}^{|I|}$ gives the average fairness score for the diffusion cascades initiated by each influencer $u$, and $\lambda[u]$ gives the expected influence spread of $u$, as the fraction of  $V$ nodes expected to be influenced by $u$ based on the norm of embedding $\|\theta\|$:
\begin{equation}\label{eq:xy}
  \lambda[u] = \frac{\|\theta_u\|_2 \times |V|}{\sum_{{u'} \in I}\|\theta_{u'}\|_2}.
\end{equation}

For the marginal gain computation at each step, the greedy algorithm works with the set of remaining, yet-to-be-influenced nodes (initially $V$ itself). At each step, it identifies / updates for each yet-to-be-selected influencer $u$ a number of $\lambda[u]$ yet-to-be-influenced nodes (lines $4$, $20$), namely those that maximize the cumulative diffusion probability from $u$, denoted $\Omega$, i.e., those associated with the top $\lambda[u]$ diffusion probabilities from $u$ (lines $5$, $21$).

Then, the influencer $u$ with the largest marginal gain -- which is a linear combination between $\Omega$ and the average fairness observed in $u$'s cascades (lines $11$, $23$) --  is selected and added to the seed set. Its top  $\lambda[u]$ influenced nodes are then removed from the yet-to-be-influenced  ones (line $17$).  In the marginal gain computation, the parameter $\alpha$ represents the aversion to unfairness, allowing us to tune the trade-off fairness vs. influence, after min-max scaling. 

As in \cite{multitask_2020}, the cumulative diffusion probability $\Omega$, marginal gain, and set consisting of $u$'s top  $\lambda[u]$ yet-to-be-influenced nodes are updated lazily, when selecting some influencer $u$ (lines $19-23$). 

The complexity of Alg. {\ref{alg:fair_greedy}} is $O(k  |I| |V| \log |V|)$. As with CELF, the algorithm benefits from (i) the fact that $|V|$ diminishes at every iteration, and (ii) much fewer influence spread evaluations than $I$ are done (lines  $19-23$). 
Further details can be found in Sec. \ref{sec:complexity}. 

\setlength{\textfloatsep}{1pt plus 1.0pt minus 2.0pt}

\begin{algorithm}[t]
		\caption{Fair-greedy selection of spread seeds}
		\label{alg:fair_greedy}
		\KwInput{ Probability diffusion matrix  $\bm{D}\in \mathbb{R}^{|I| \times |V|}$, cas\-cade ba\-sed fairness vector $\bm{F} \in \mathbb{R}^{|I|}$, expected fair spread vector $\lambda \in \mathbb{R}^{|I|}$, seed set size $k$, parameter $\alpha$}
		\KwOutput{seed set $\mathcal{S}$}
		
		$Q \gets [\,] , \,  \mathcal{S} \gets \emptyset , \,   L \gets [0: |V|]$ \\
		\For{$s=0$; $s<|I|$; $s++$}
		{
			$R \gets\text{argsort}(D[s,L])[0:\lambda[s]]$ \\
			$\Omega \gets \text{sum}(D[s,R])$ \\
			$Q.\text{append}([s,\Omega,F[s],R,0])$
		}
		$\Omega_{ub} \gets \text{max}(Q,1), \, \Omega_{lb} \gets \text{min}(Q,1)$\\
		$F_{ub} \gets \text{max}(F), \, F_{lb} \gets \text{min}(F)$\\
		\For{$q \in Q $}
		{
			$q[1] \gets F_{lb} + (q[1]-\Omega_{lb})/(\Omega_{ub}-\Omega_{lb})\times (F_{ub}-F_{lb})$ \\
		}	
	$Q \gets \text{sort}(Q,\text{lambda $q \in Q$:}~(1-\alpha) \times q[1] +\alpha \times q[2])$ \\
			
  \While{$|\mathcal{S}| < k$}
		{
		$q \gets Q[0] $ \\
			\If{$q[4] == |\mathcal{S}| $}
			{
				$\mathcal{S}\text{.add}(q[0])$ \\
				$R \gets q[3]$\\
				$L \gets L-R$ \\
				$Q.\text{delete}(q)$\\
			}
		\Else
			{	$R \gets \text{argsort}(D[q(0),L])[0:\lambda[q[0]]]$ \\
			$\Omega \gets \text{sum}(D[q[0],R])$ \\

			$Q[0] \gets [q[0],\Omega,F[q[0]],R,|\mathcal{S}|]$ \\
			$Q \gets \text{sort}(Q,\text{lambda $q\in Q$:}~(1-\alpha) \times q[1] +\alpha \times q[2])$ \\
			}
		}
		\Return $\mathcal{S}$
	\end{algorithm}

%% file: experiments.tex
\label{sec:experiments}
We evaluated  the proposed algorithms on two datasets. Weibo is the largest Chinese microblogging platform, and the Weibo dataset \cite{weibopaper} is a publicly available one built for information diffusion studies. It includes all the facets of interest for our work, namely a diffusion graph, the information cascades, and  user profiles. Despite our best effort to find a similar, real-world and publicly available dataset, we could not identify one containing  these data facets, in particular ground-truth cascades and user profiles. Therefore, to reduce the risk of validation biases, we built a hybrid real-synthetic dataset, based on the publicly available Digg2009 dataset \cite{https://doi.org/10.48550/arxiv.1202.3162}. Digg is a platform where users follow others, (re)post stories, and vote on posts. This dataset contains all the necessary facets, except for the user profiles. Hence we complemented it by synthetically generated user profiles (sensitive attributes), which overall fit  publicly-known  distributions in Digg. The main data statistics are in Table.{\ref{tab:tab_datasets}}. The experiments were done on a server Ubuntu 20.04.4 LTS Intel   CPU E5-2620 v2 @ 2.10GHz, with
 188GB RAM and Tesla k40m GPU.

\begin{table}[t!]
  \caption{Datasets description.}
  \label{tab:tab_datasets}
  \begin{tabular}{lll}
    \toprule
         & {\itshape Weibo} &{\itshape Digg} \\
    \midrule
          influencers & 19,698 & 532\\
          influencees & 1,170,688 & 336,225 \\
          social links & 225,877,632& 2,617,993 \\
          posts &62,087 & 3,553 \\
          median size of cascades & 50 & 528\\
          maximal  size  of  cascades & 35986 & 24099\\
  \bottomrule
\end{tabular}
\end{table}

\begin{figure*}[t]
  \centering
  \includegraphics[width=0.88\linewidth]{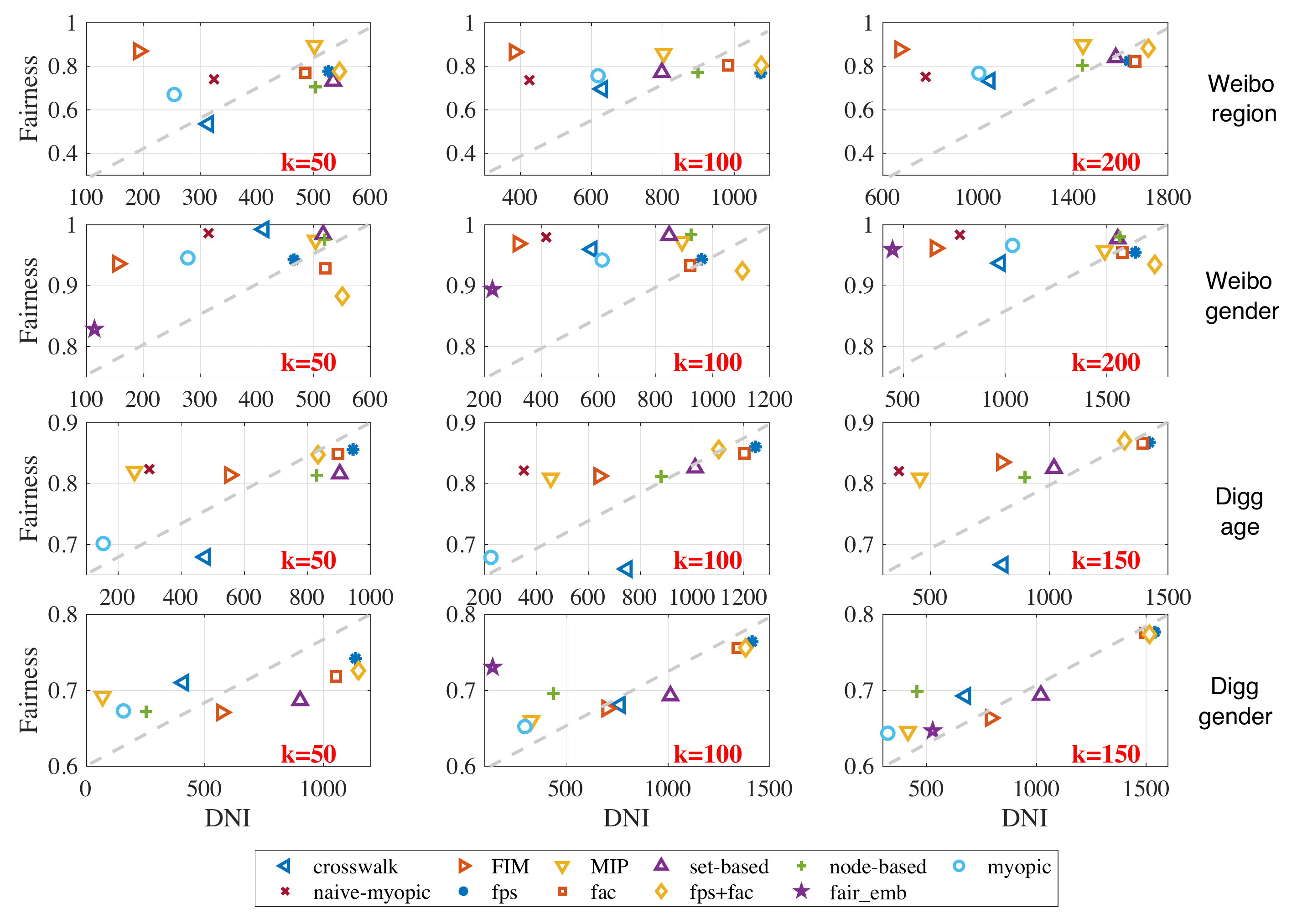}
  \caption{Validation on the sampled datasets from Weibo and Digg (the closer to the upper-right corner the better).}
  \label{fig:sampled_comparison}
\end{figure*}

\subsection{Baseline methods and setup}
\label{sec:baselines}
The state-of-the-art methods on fair IM based on graph algorithms \cite{DBLP:conf/aaai/RahmattalabiJLV21,DBLP:conf/www/FarnadiBG20, DBLP:conf/aaai/BeckerDGG21,gaps2019,groupinfluencemax2019} cannot scale to large diffusion scenarios with thousands-to-millions of users. In contrast, by using the ground-truth diffusion cascades instead of the social connectivity, and by learning node representations based on them,  our models are able to deal with social graphs and information diffusion at large scale. The same scalability limits are exhibited by the two  state-of-the-art methods on fair IM that do rely on node representation \cite{adversialgraph2020,khajehnejad2022crosswalk}, yet learned only from the social connectivity. For all these methods, their respective empirical validation results,  performed on extremely small datasets, are clear indicators of their high computation cost and thus limited applicability.  Therefore, for a comparison with  these methods, 
we sampled smaller representative Weibo / Digg datasets, with thousands of nodes, on which their running time remained nevertheless quite high (mostly order of hours).  

 However,  we also experimented with the complete datasets, com\-paring with IM models that are (i) applicable to large social graphs, (ii) exploit  diffusion cascades (as we do),  but (iii) are fairness agnostic. This enabled us to showcase that the fairness achieved by our models, at realistic scale, does not come at the expense of spread.  

\vspace{-1.5mm}
\paragraph{Fair IM models (on sampled datasets).} 
\textbf{Crosswalk} \cite{khajehnejad2022crosswalk} is a random-walk based graph representation method,  which enhances fairness by re-weigh\-ting the edges between nodes from different groups. We initialize the edge weights based on the in-out degree and, with the resulting embeddings, the potential seeds are selected via clustering. 
\textbf{FairEmbedding} \cite{adversialgraph2020} uses an adversarial network for graph embedding, with a discriminator to discern the sensitive attributes.  By design, it is only applicable for attributes with binary discrimination, so its experiments are only carried out on \emph{gender}. The potential seeds are also selected by clustering based on the resulting embeddings. The graph algorithms denoted \textbf{Node-based} / \textbf{Set-based} from \cite{DBLP:conf/aaai/BeckerDGG21} use probabilistic strategies over nodes and communities to select seeds for fair IM under the maximin formulation.  Their edge weights are also initialized based on in-out degrees.  \textbf{FIM} \cite{DBLP:conf/aaai/RahmattalabiJLV21} proposes a framework based on social-welfare theory to mitigate the fairness vs. spread trade-off, with a single inequality-aversion parameter; it employs the notion of fairness under equity (our case) when the inequality-aversion parameter is set to  $0$. \textbf{MIP} \cite{DBLP:conf/www/FarnadiBG20} achieves fairness under equity in IM by adding linear constraints or formulating the objective function as an integer program.  
Finally, \textbf{Myopic} / \textbf{Naive-myopic} \cite{gaps2019} are heuristic graph algorithms that select  seeds based on maximin group fairness; the former iteratively chooses seeds based on minimum probability of reaching groups, while the later choose the seeds at once.

We also compare with \textbf{FPS+FAC}, a hybrid solution where we take the down-sampling strategy of FPS as input, while learning the fairness score as in FAC, in the output of network, in order to investigate their potential to jointly improve performance.


\vspace{-1.5mm}
\paragraph{Fairness-agnostic IM methods (on complete datasets).} Since the fair IM models cannot be evaluated on the complete datasets, we look instead at fairness-agnostic IM methods using the diffusion cascades  to identify and select the spread seeds. \textbf{IMINFECTOR} \cite{multitask_2020} designs a multi-task graph neural network,  which takes as input the cascades to generate node embeddings, by which both diffusion probabilities (between influencers and influencees) and cascade sizes can be predicted.    A seed set is selected based on these, by a CELF-like greedy algorithm. Our work draws inspiration from  \cite{multitask_2020} in both the way node representations are learned from cascades, and in the subsequent greedy algorithm for fair IM.  \textbf{CELFIE} \cite{panagopoulos2020influence} is a model preceding \textbf{IMINFECTOR}, where the network is trained on cascades to learn the node embeddings, but not the spread aptitude. Finally, we consider a data-based heuristic,  \textbf{AvgCascades} \cite{bakshy2011everyone}, which ranks the seeds based on the average size of their diffusion cascades. 

\vspace{-1.5mm}
\paragraph{Setup.}
 We split the cascades into train (60\%), validation (20\%), and test (20\%) subsets,  by occurrence time.  
  For the deep neural networks, the embedding dimension $|E|$ is set to 50, the models are trained for 10 epochs, with a learning rate of 0.1. The original context is over-sampled by $\eta=120\%$. In FPS, to avoid sparsity and to ensure that most influencers can be used in the training stage, we applied a threshold on the maximum penalization exerted on those having low fairness, such that at least $3$ cascades are kept for training.  In the fair-greedy seed selection, the parameter $\alpha$ indicating the aversion to unfairness is by default set to $0.2$. 
 As the sensitive attributes, in  Weibo we consider  \emph{gender} and \emph{region}, while in Digg we consider \emph{gender} and \emph{age}. The \emph{gender} distribution is around 53\% \emph{female} and 47\% \emph{male} in Weibo, respectively  35\% \emph{female} and 65\% \emph{male} in Digg.  
  (Sec. \ref{sec:distributions} details the distribution of Weibo \emph{region} and Digg \emph{age}.)  
Performance is  evaluated on fairness and spread, based on the test cascades. Spread is given by the DNI metric (Distinct Nodes Influenced) \cite{scalableinfluence2013},  as a union over the nodes that participate in the test cascades initiated by the selected seeds.

\subsection{Validation on sampled datasets}
\label{sec:sample_graph}
 The results are presented in Fig. {\ref{fig:sampled_comparison}} where, for each sub-plotting, the $x$ axis represents DNI and the $y$ axis represents fairness. To give a clearer view on the differences in performance, we only keep the fairness region where most models lie, which means that those with significantly worse performance may be omitted in some plots.  

We can see that, as the size of the seed set increases, both the number of influenced nodes and the fairness score goes up for almost all the models. Our models generally outperform the other baselines, especially on the influence dimension (DNI). As for fairness, our models remain consistently in the top-tier, except when applied to Weibo \emph{gender}, a case which we will further explore next. 

\begin{figure}[t!]
  \centering
  \includegraphics[width=0.86\linewidth]{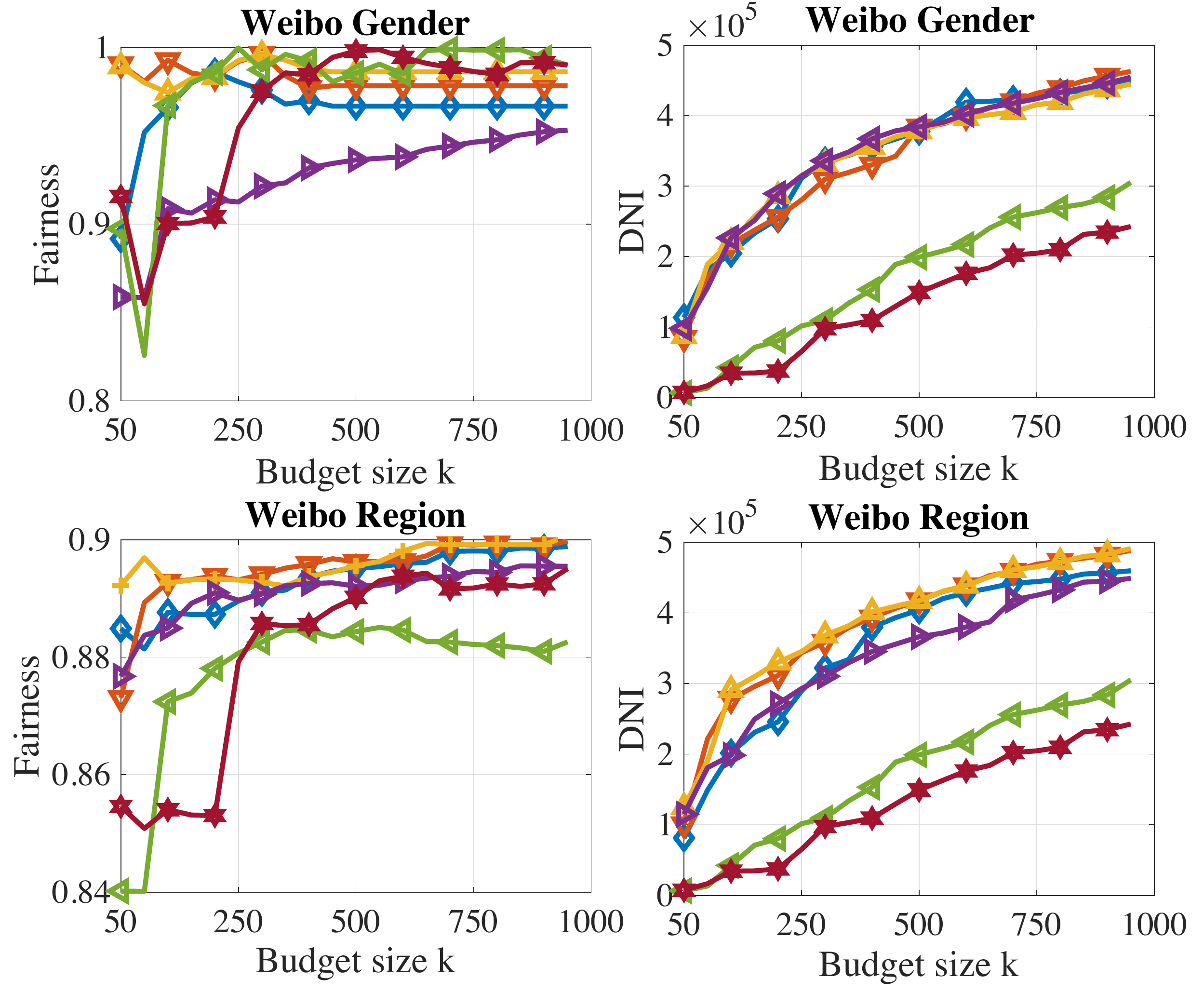}
  \includegraphics[width=0.86\linewidth]{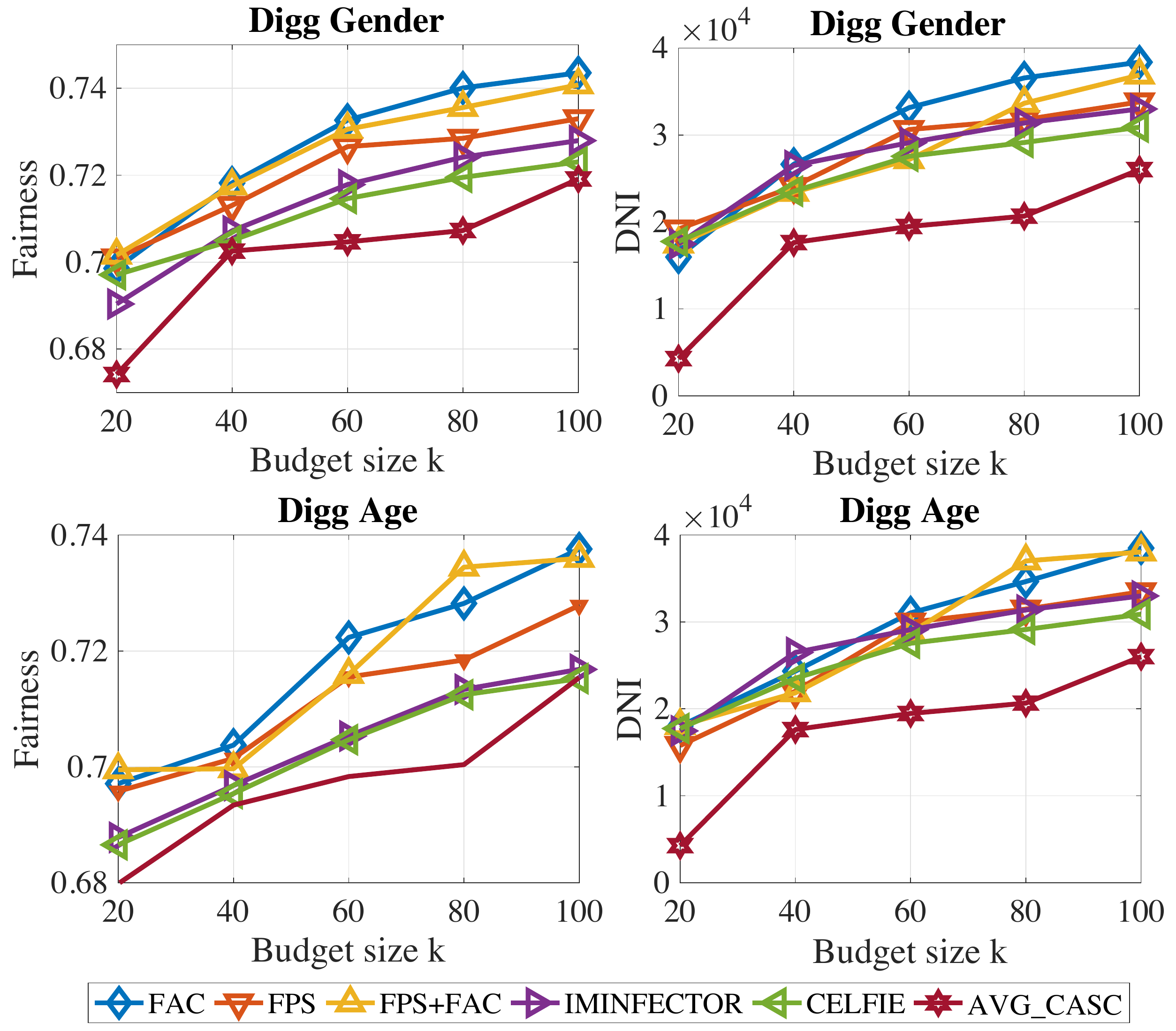}
  \caption{Validation on the complete datasets.}
  \label{fig:complet_weibo_digg}
\end{figure}

Among the baselines, \textbf{Set-based} and \textbf{Node-based} generally outperform the other baselines, with stable performance on fairness and influence, proving the effectiveness of selecting seeds with  randomization. \textbf{FIM} and \textbf{MIP} perform generally well on fairness, but often fail in maximizing the spread. This may be due to the fact that their focus is on fairness under various flavors (among which the one we focus on, equity), by a generic formal and algorithmic solution,  which may come at the expense of influence spread.

As to the embedding-based methods, \textbf{Crosswalk} and \textbf{FairEmbedding}, their algorithms for selecting the seeds are clustering based, 
leading to rather unstable performance along both dimensions of interest.  In that regard, our fair-greedy selection algorithm after the embedding process proves to be much more effective. 

 Among the  models we propose, FPS performs relatively worse in Weibo, while in Digg they are performing almost equally well. As the down-sampling strategy as a penalization on unfair influencers in FPS is independent from the fairness score learned in FAC,  we do not necessarily benefit from the combination of these two methods.

\subsection{Validation on the complete datasets}
\label{sec:complet_graph}

\eat{
\begin{figure}[t!]
\vspace{-3.5mm}
  \centering
  \includegraphics[width=1\linewidth]{img/digg_complet_2.pdf}
  \vspace{-2mm}
  \caption{Validation on complete  Digg dataset.}
  \label{fig:complet_digg}
\vspace{-1mm}
\end{figure}
}

As the most related state-of-the-art methods are not scalable on the realistic datasets, we further validate our solutions by comparison with generic, cascade-based IM models,  in order to showcase their scalability and effectiveness (Fig. \ref{fig:complet_weibo_digg}). As expected,  our models outperform the baselines on fairness, in both Weibo and Digg. Once again, and probably even more so now, since we have large casca
%
des,   the case of Weibo \emph{gender} is an exception, since large spread translates automatically and ``effortless''  to high fairness. 
The performance of our methods on influence remains consistently in the top-tier, proving that the fairness gains do not come at the  expense of influence, and that spread maximization can remain the top objective. 

It is worth noting that \textbf{IMINFECTOR} not only has good (top-tier) spread performance, but also does reasonably well on fairness. Our explanation is that, at large scale and as a feature of our datasets, the larger the audience an influencer may have, the closer that audience's  distribution on sensitive attributes will be to the one of the entire population. Therefore, under the equity formulation, high fairness becomes well-correlate to high spread. 




\subsection{Validation on an unbalanced Weibo dataset}
\label{sec:unbalanced_flipping}
Recall from the results of Fig. {\ref{fig:sampled_comparison}}  -- sampled Weibo data,  \emph{gender} fairness --  that  although we outperform the other baselines on spread, the differences on fairness between models are limited. \eat{A similar observation can be made for Weibo \emph{gender} in Fig. \ref{fig:complet_weibo_digg}, as large spread implies high fairness.}
We can also note that the fairness scores on Weibo \emph{gender} are higher than for other attributes. Our take on this is that, since \emph{gender} is already quite balanced, the diffusions follow that same distribution and thus only minor differences in  fairness performance are to be expected. 
 In Fig. {\ref{fig:sampled_comparison}}, the plots fall within a thin y-region close to the $1$ bar, while being scattered on the x-axis, which means all models can do quite well on fairness, while exhibiting  differences on spread. 

\begin{figure}[t!]
  \centering
  \includegraphics[width=0.94\linewidth]{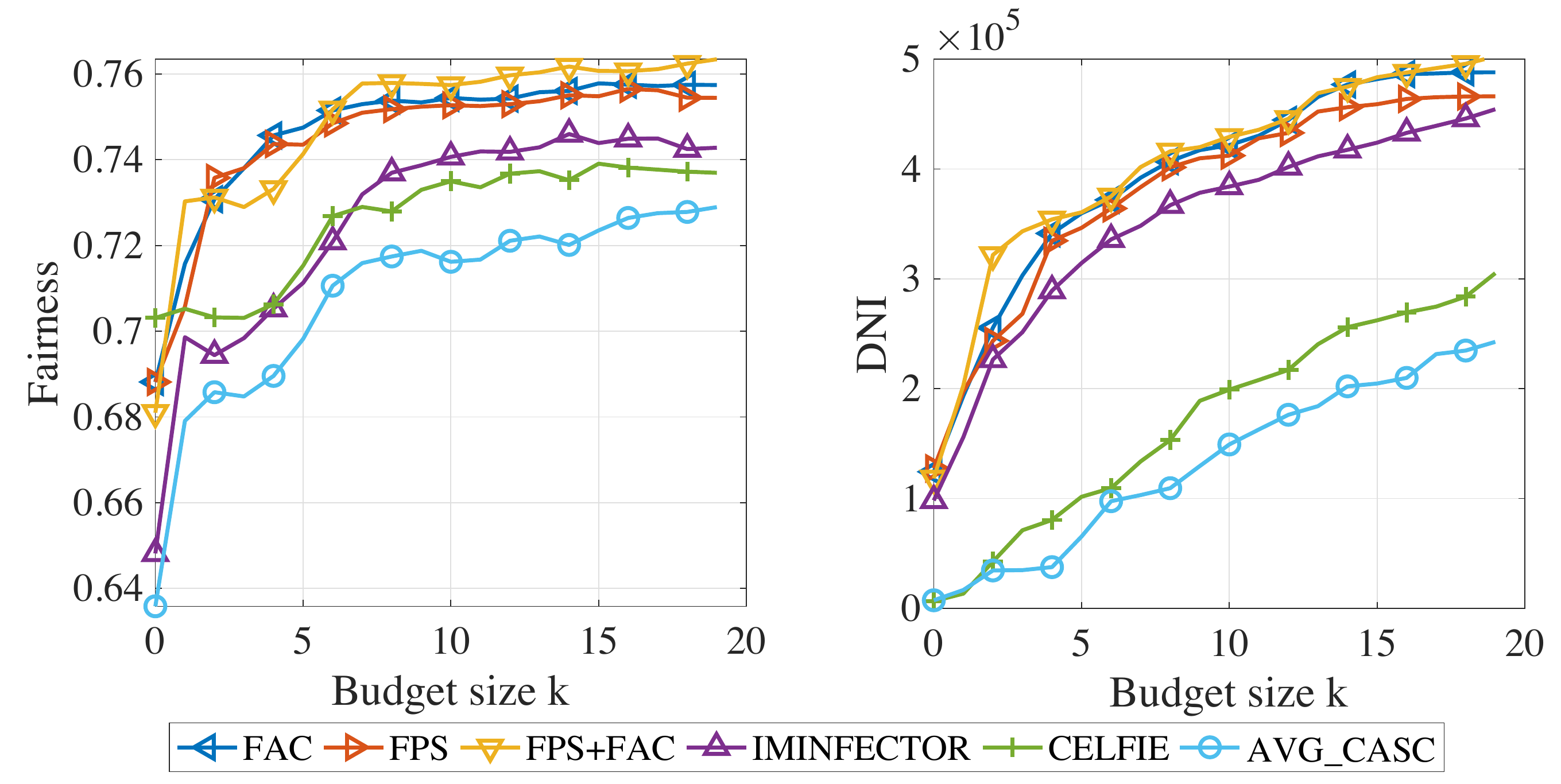}
  \caption{Unbalanced-\emph{gender} complete Weibo dataset.}
  \label{fig:flipped_weibo_complet}
\end{figure}

\begin{figure}[b!]
  \centering
  \includegraphics[width=0.91\linewidth]{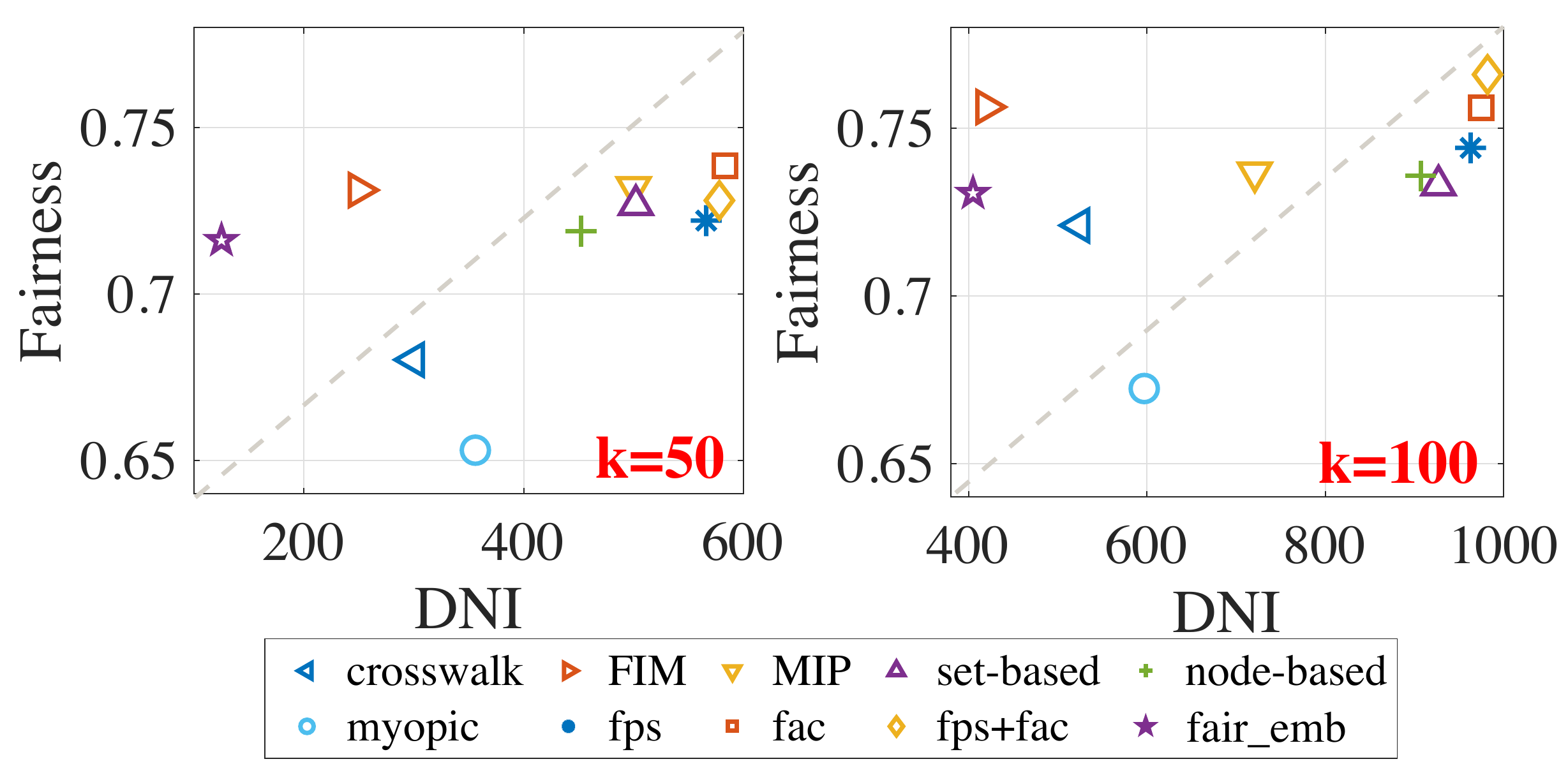}
  \caption{Validation on the unbalanced-\emph{gender} Weibo sample.}
  \label{fig:synthetic_weibo}
\end{figure}

To further investigate this aspect, in Weibo (complete and sampled), we did a set of experiments where we 
``flip'' \emph{gender} profile values in the cascades of some randomly selected influencers, for randomly selected participants thereof. The rationale was to obtain some unbalanced influencers, i.e., infuencers with unfair cascades, while reaching an overall  \emph{male / female}  ratio of roughly $25\%-75\%$.  
The distribution of the fairness score of influencers,  before  / after flipping, is relegated to Sec. \ref{sec:flipping} (Fig.{\ref{fig:distribution_flipping}}). 

In Fig. \ref{fig:flipped_weibo_complet}, we present  the comparison with the fairness-agnostic methods.  We can observe now that when we decorrelate high spread potential and high fairness, the fairness gap between our methods and the baselines (in particular \textbf{IMINFECTOR}) becomes larger.

\begin{figure}[t!]
  \centering
  \includegraphics[width=0.87\linewidth]{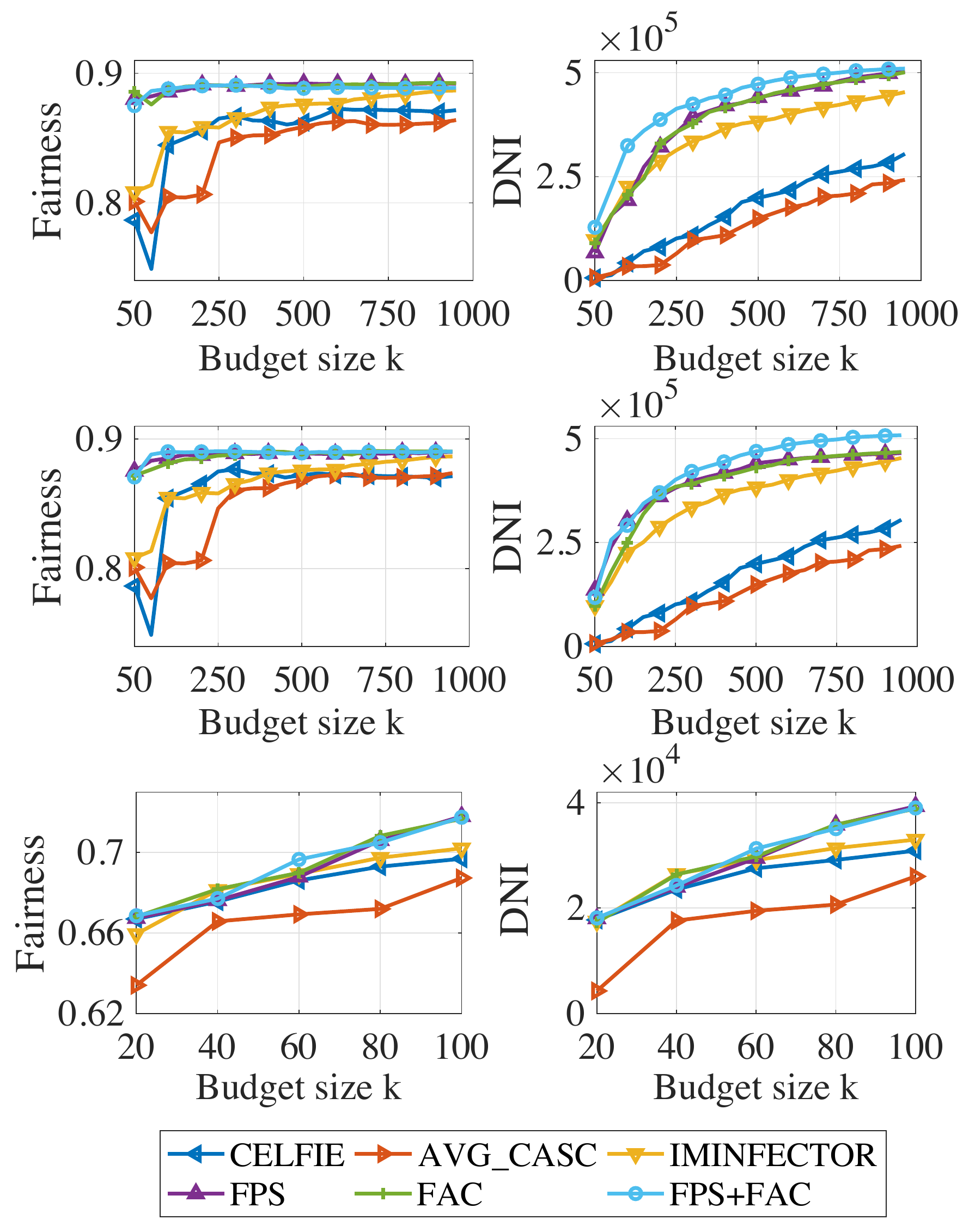}
  \caption{Evaluation for combinations of attributes (top row: Weibo, mid-row: flipped-\emph{gender} Weibo, bottom row: Digg).}
  \label{fig:multiple_attributes}
\end{figure}


In Fig. {\ref{fig:synthetic_weibo}}, we have the evaluation on the sampled data. We can notice the performance shifts, as our models have now a clear advantage when facing the risk of large yet unfair propagations, which also indicates they are more robust to such unbalancedness. 

\subsection{Combinations of sensitive attributes}
\label{sec:combinations}
Recall that our fairness model applies not only to individual sensitive attributes, but also to combinations thereof. 
 We present in this section our evaluation results on combinations of attributes: \emph{gender\_region} in Weibo (complete), with or without \emph{gender} flipping, and \emph{gender\_age} in Digg (complete).   In Weibo, with 2 \emph{gender}  categories and 36 \emph{region} ones, we obtain a combined sensitive attribute having 72 categories (e.g., ``women in region 1''). Similarly, in Digg, 
 we have a combined sensitive attribute with 12 categories.

The results are given in Fig.\ref{fig:multiple_attributes}. We can notice that all our models show once again clear advantages over the baselines on both  fairness and influence. Interestingly, the results on the original Weibo data and on the one with flipped \emph{gender} are quite similar now, as the ``effortless fairness'' phenomenon caused by the balanced \emph{gender} distribution is now attenuated by aggregation with \emph{region}.

Finally, we considered also the following question: \emph{can we train our models on individual sensitive attributes, and then use the resulting embeddings for fair IM on combinations of attributes}? Note that this could be very beneficial, allowing us to avoid a costly training phase whenever  many (or all) combinations of sensitive attributes may arise in fair IM queries.  We describe some initial results on this aspect in Section \ref{aggregative-concatenation}.

%% file: conclusion.tex
\label{sec:conclusion}
We revisit the problem of  influence maximization  with fairness.  We propose two deep-learning based algorithms for extracting node representations from a history of diffusion cascades.  
Unlike the prior works, the proposed algorithms are generic -- applicable to arbitrary sets of sensitive attributes -- and scalable. Our embedding algorithms take into account both the spread aptitude and the fairness of influencers, and use the resulting embeddings to greedily select the optimal spread seeds. The experimental results, performed on both real-world and synthetic data, show that our methods outperform the state-of-the-art methods for fair IM (with limited scalability) or for fairness-agnostic IM based on cascades.  
Future work will investigate how the training could be done efficiently in a single stage for all the sensitive attributes and how learned representations pertaining to individual sensitive attributes could be ``transferred'' to combinations thereof.  


%% file: sample-authordraft.bbl

\begin{thebibliography}{30}


\ifx \showCODEN    \undefined \def \showCODEN     #1{\unskip}     \fi
\ifx \showDOI      \undefined \def \showDOI       #1{#1}\fi
\ifx \showISBNx    \undefined \def \showISBNx     #1{\unskip}     \fi
\ifx \showISBNxiii \undefined \def \showISBNxiii  #1{\unskip}     \fi
\ifx \showISSN     \undefined \def \showISSN      #1{\unskip}     \fi
\ifx \showLCCN     \undefined \def \showLCCN      #1{\unskip}     \fi
\ifx \shownote     \undefined \def \shownote      #1{#1}          \fi
\ifx \showarticletitle \undefined \def \showarticletitle #1{#1}   \fi
\ifx \showURL      \undefined \def \showURL       {\relax}        \fi
\providecommand\bibfield[2]{#2}
\providecommand\bibinfo[2]{#2}
\providecommand\natexlab[1]{#1}
\providecommand\showeprint[2][]{arXiv:#2}

\bibitem[Ali et~al\mbox{.}(2022)]%
        {onthefairness2019}
\bibfield{author}{\bibinfo{person}{J. Ali}, \bibinfo{person}{M. Babaei},
  \bibinfo{person}{A. Chakraborty}, \bibinfo{person}{B. Mirzasoleiman},
  \bibinfo{person}{K. Gummadi}, {and} \bibinfo{person}{A. Singla}.}
  \bibinfo{year}{2022}\natexlab{}.
\newblock \showarticletitle{On the Fairness of Time-Critical Influence
  Maximization in Social Networks}.
\newblock \bibinfo{journal}{\emph{IEEE Transactions on Knowledge \& Data
  Engineering}} \bibinfo{number}{01} (\bibinfo{year}{2022}),
  \bibinfo{pages}{1--1}.
\newblock
\showISSN{1558-2191}


\bibitem[Aral and Dhillon(2018)]%
        {socialinfluencemax2018}
\bibfield{author}{\bibinfo{person}{Sinan Aral} {and}
  \bibinfo{person}{Paramveer~S. Dhillon}.} \bibinfo{year}{2018}\natexlab{}.
\newblock \showarticletitle{Social influence maximization under empirical
  influence models}.
\newblock \bibinfo{journal}{\emph{Nature Human Behaviour}}  \bibinfo{volume}{2}
  (\bibinfo{year}{2018}), \bibinfo{pages}{375--382}.
\newblock


\bibitem[Arora et~al\mbox{.}(2017)]%
        {DBLP:conf/sigmod/AroraGR17}
\bibfield{author}{\bibinfo{person}{Akhil Arora}, \bibinfo{person}{Sainyam
  Galhotra}, {and} \bibinfo{person}{Sayan Ranu}.}
  \bibinfo{year}{2017}\natexlab{}.
\newblock \showarticletitle{Debunking the Myths of Influence Maximization: An
  In-Depth Benchmarking Study}. In \bibinfo{booktitle}{\emph{SIGMOD}}.
\newblock


\bibitem[Arora et~al\mbox{.}(2019)]%
        {DBLP:conf/edbt/0001GR19}
\bibfield{author}{\bibinfo{person}{Akhil Arora}, \bibinfo{person}{Sainyam
  Galhotra}, {and} \bibinfo{person}{Sayan Ranu}.}
  \bibinfo{year}{2019}\natexlab{}.
\newblock \showarticletitle{Influence Maximization Revisited: The State of the
  Art and the Gaps that Remain}. In \bibinfo{booktitle}{\emph{EDBT}}.
\newblock


\bibitem[Bakshy et~al\mbox{.}(2011)]%
        {bakshy2011everyone}
\bibfield{author}{\bibinfo{person}{Eytan Bakshy}, \bibinfo{person}{Jake~M
  Hofman}, \bibinfo{person}{Winter~A Mason}, {and} \bibinfo{person}{Duncan~J
  Watts}.} \bibinfo{year}{2011}\natexlab{}.
\newblock \showarticletitle{Everyone's an influencer: quantifying influence on
  twitter}. In \bibinfo{booktitle}{\emph{WSDM}}.
\newblock


\bibitem[Becker et~al\mbox{.}(2021)]%
        {DBLP:conf/aaai/BeckerDGG21}
\bibfield{author}{\bibinfo{person}{Ruben Becker}, \bibinfo{person}{Gianlorenzo
  D'Angelo}, \bibinfo{person}{Sajjad Ghobadi}, {and} \bibinfo{person}{Hugo
  Gilbert}.} \bibinfo{year}{2021}\natexlab{}.
\newblock \showarticletitle{Fairness in Influence Maximization through
  Randomization}. In \bibinfo{booktitle}{\emph{AAAI}}.
\newblock


\bibitem[Bourigault et~al\mbox{.}(2016)]%
        {representationlearning2016}
\bibfield{author}{\bibinfo{person}{Simon Bourigault}, \bibinfo{person}{Sylvain
  Lamprier}, {and} \bibinfo{person}{Patrick Gallinari}.}
  \bibinfo{year}{2016}\natexlab{}.
\newblock \showarticletitle{Representation Learning for Information Diffusion
  through Social Networks: An Embedded Cascade Model}. In
  \bibinfo{booktitle}{\emph{WSDM}}.
\newblock


\bibitem[Brown and Fiorella(2013)]%
        {InfluencerMarketing}
\bibfield{author}{\bibinfo{person}{Danny Brown} {and} \bibinfo{person}{Sam
  Fiorella}.} \bibinfo{year}{2013}\natexlab{}.
\newblock \bibinfo{booktitle}{\emph{{Influence Marketing: How to Create,
  Manage, and Measure Brand Influencers in Social MediaMarketing}}}.
\newblock \bibinfo{publisher}{Que Pub}.
\newblock


\bibitem[Du et~al\mbox{.}(2013)]%
        {scalableinfluence2013}
\bibfield{author}{\bibinfo{person}{Nan Du}, \bibinfo{person}{Le Song},
  \bibinfo{person}{Manuel Gomez~Rodriguez}, {and} \bibinfo{person}{Hongyuan
  Zha}.} \bibinfo{year}{2013}\natexlab{}.
\newblock \showarticletitle{Scalable Influence Estimation in Continuous-Time
  Diffusion Networks}. In \bibinfo{booktitle}{\emph{NIPS}}.
\newblock


\bibitem[Farnadi et~al\mbox{.}(2020)]%
        {DBLP:conf/www/FarnadiBG20}
\bibfield{author}{\bibinfo{person}{Golnoosh Farnadi}, \bibinfo{person}{Behrouz
  Babaki}, {and} \bibinfo{person}{Michel Gendreau}.}
  \bibinfo{year}{2020}\natexlab{}.
\newblock \showarticletitle{A Unifying Framework for Fairness-Aware Influence
  Maximization}. In \bibinfo{booktitle}{\emph{The Web Conference}}.
\newblock


\bibitem[Feng et~al\mbox{.}(2018)]%
        {inf2vec}
\bibfield{author}{\bibinfo{person}{Shanshan Feng}, \bibinfo{person}{Gao Cong},
  \bibinfo{person}{Arijit Khan}, \bibinfo{person}{Xiucheng Li},
  \bibinfo{person}{Yong Liu}, {and} \bibinfo{person}{Yeow~Meng Chee}.}
  \bibinfo{year}{2018}\natexlab{}.
\newblock \showarticletitle{Inf2vec: Latent Representation Model for Social
  Influence Embedding}. In \bibinfo{booktitle}{\emph{ICDE}}.
\newblock


\bibitem[Fish et~al\mbox{.}(2019)]%
        {gaps2019}
\bibfield{author}{\bibinfo{person}{Benjamin Fish}, \bibinfo{person}{Ashkan
  Bashardoust}, \bibinfo{person}{Danah Boyd}, \bibinfo{person}{Sorelle
  Friedler}, \bibinfo{person}{Carlos Scheidegger}, {and}
  \bibinfo{person}{Suresh Venkatasubramanian}.}
  \bibinfo{year}{2019}\natexlab{}.
\newblock \showarticletitle{Gaps in Information Access in Social Networks}. In
  \bibinfo{booktitle}{\emph{WWW}}.
\newblock


\bibitem[Gershtein et~al\mbox{.}(2021)]%
        {DBLP:conf/edbt/GershteinMY21}
\bibfield{author}{\bibinfo{person}{Shay Gershtein}, \bibinfo{person}{Tova
  Milo}, {and} \bibinfo{person}{Brit Youngmann}.}
  \bibinfo{year}{2021}\natexlab{}.
\newblock \showarticletitle{Multi-Objective Influence Maximization}. In
  \bibinfo{booktitle}{\emph{EDBT}}, \bibfield{editor}{\bibinfo{person}{Yannis
  Velegrakis}, \bibinfo{person}{Demetris Zeinalipour{-}Yazti},
  \bibinfo{person}{Panos~K. Chrysanthis}, {and} \bibinfo{person}{Francesco
  Guerra}} (Eds.).
\newblock


\bibitem[Grover and Leskovec(2016)]%
        {grover2016node2vec}
\bibfield{author}{\bibinfo{person}{Aditya Grover} {and} \bibinfo{person}{Jure
  Leskovec}.} \bibinfo{year}{2016}\natexlab{}.
\newblock \showarticletitle{node2vec: Scalable feature learning for networks}.
  In \bibinfo{booktitle}{\emph{SIGKDD}}.
\newblock


\bibitem[Kempe et~al\mbox{.}(2003)]%
        {kempe2003maximizing}
\bibfield{author}{\bibinfo{person}{David Kempe}, \bibinfo{person}{Jon
  Kleinberg}, {and} \bibinfo{person}{{\'E}va Tardos}.}
  \bibinfo{year}{2003}\natexlab{}.
\newblock \showarticletitle{Maximizing the spread of influence through a social
  network}. In \bibinfo{booktitle}{\emph{ACM SIGKDD}}.
\newblock


\bibitem[Khajehnejad et~al\mbox{.}(2022)]%
        {khajehnejad2022crosswalk}
\bibfield{author}{\bibinfo{person}{Ahmad Khajehnejad}, \bibinfo{person}{Moein
  Khajehnejad}, \bibinfo{person}{Mahmoudreza Babaei},
  \bibinfo{person}{Krishna~P Gummadi}, \bibinfo{person}{Adrian Weller}, {and}
  \bibinfo{person}{Baharan Mirzasoleiman}.} \bibinfo{year}{2022}\natexlab{}.
\newblock \showarticletitle{CrossWalk: fairness-enhanced node representation
  learning}. In \bibinfo{booktitle}{\emph{AAAI}}.
\newblock


\bibitem[Khajehnejad et~al\mbox{.}(2020)]%
        {adversialgraph2020}
\bibfield{author}{\bibinfo{person}{Moein Khajehnejad}, \bibinfo{person}{Ahmad
  Asgharian~Rezaei}, \bibinfo{person}{Mahmoudreza Babaei},
  \bibinfo{person}{Jessica Hoffmann}, \bibinfo{person}{Mahdi Jalili}, {and}
  \bibinfo{person}{Adrian Weller}.} \bibinfo{year}{2020}\natexlab{}.
\newblock \showarticletitle{Adversarial Graph Embeddings for Fair Influence
  Maximization over Social Networks}. In \bibinfo{booktitle}{\emph{IJCAI}}.
\newblock


\bibitem[Lerman et~al\mbox{.}(2012)]%
        {https://doi.org/10.48550/arxiv.1202.3162}
\bibfield{author}{\bibinfo{person}{Kristina Lerman}, \bibinfo{person}{Rumi
  Ghosh}, {and} \bibinfo{person}{Tawan Surachawala}.}
  \bibinfo{year}{2012}\natexlab{}.
\newblock \bibinfo{title}{Social Contagion: An Empirical Study of Information
  Spread on Digg and Twitter Follower Graphs}.
\newblock
\newblock
\urldef\tempurl%
\url{https://doi.org/10.48550/ARXIV.1202.3162}
\showDOI{\tempurl}


\bibitem[Leskovec et~al\mbox{.}(2007)]%
        {DBLP:conf/kdd/LeskovecKGFVG07}
\bibfield{author}{\bibinfo{person}{Jure Leskovec}, \bibinfo{person}{Andreas
  Krause}, \bibinfo{person}{Carlos Guestrin}, \bibinfo{person}{Christos
  Faloutsos}, \bibinfo{person}{Jeanne~M. VanBriesen}, {and}
  \bibinfo{person}{Natalie~S. Glance}.} \bibinfo{year}{2007}\natexlab{}.
\newblock \showarticletitle{Cost-effective outbreak detection in networks}. In
  \bibinfo{booktitle}{\emph{ACM SIGKDD}}.
\newblock


\bibitem[{Li} et~al\mbox{.}(2018)]%
        {8295265}
\bibfield{author}{\bibinfo{person}{Y. {Li}}, \bibinfo{person}{J. {Fan}},
  \bibinfo{person}{Y. {Wang}}, {and} \bibinfo{person}{K. {Tan}}.}
  \bibinfo{year}{2018}\natexlab{}.
\newblock \showarticletitle{Influence Maximization on Social Graphs: A Survey}.
\newblock \bibinfo{journal}{\emph{IEEE Transactions on Knowledge and Data
  Engineering}} \bibinfo{volume}{30}, \bibinfo{number}{10}
  (\bibinfo{year}{2018}), \bibinfo{pages}{1852--1872}.
\newblock
\urldef\tempurl%
\url{https://doi.org/10.1109/TKDE.2018.2807843}
\showDOI{\tempurl}


\bibitem[Lin et~al\mbox{.}(2020)]%
        {DBLP:conf/wsdm/LinLL20}
\bibfield{author}{\bibinfo{person}{Mingkai Lin}, \bibinfo{person}{Wenzhong Li},
  {and} \bibinfo{person}{Sanglu Lu}.} \bibinfo{year}{2020}\natexlab{}.
\newblock \showarticletitle{Balanced Influence Maximization in Attributed
  Social Network Based on Sampling}. In \bibinfo{booktitle}{\emph{{WSDM} '20:
  The Thirteenth {ACM} International Conference on Web Search and Data Mining,
  Houston, TX, USA, February 3-7, 2020}},
  \bibfield{editor}{\bibinfo{person}{James Caverlee},
  \bibinfo{person}{Xia~(Ben) Hu}, \bibinfo{person}{Mounia Lalmas}, {and}
  \bibinfo{person}{Wei Wang}} (Eds.). \bibinfo{pages}{375--383}.
\newblock


\bibitem[Panagopoulos et~al\mbox{.}(2020b)]%
        {multitask_2020}
\bibfield{author}{\bibinfo{person}{George Panagopoulos},
  \bibinfo{person}{Fragkiskos Malliaros}, {and} \bibinfo{person}{M
  Vazirgiannis}.} \bibinfo{year}{2020}\natexlab{b}.
\newblock \showarticletitle{Multi-task Learning for Influence Estimation and
  Maximization}.
\newblock \bibinfo{journal}{\emph{IEEE Transactions on Knowledge and Data
  Engineering}} (\bibinfo{year}{2020}).
\newblock


\bibitem[Panagopoulos et~al\mbox{.}(2020a)]%
        {panagopoulos2020influence}
\bibfield{author}{\bibinfo{person}{George Panagopoulos},
  \bibinfo{person}{Fragkiskos~D Malliaros}, {and} \bibinfo{person}{Michalis
  Vazirgianis}.} \bibinfo{year}{2020}\natexlab{a}.
\newblock \showarticletitle{Influence maximization using influence and
  susceptibility embeddings}. In \bibinfo{booktitle}{\emph{Proceedings of the
  International AAAI Conference on Web and Social Media}},
  Vol.~\bibinfo{volume}{14}.
\newblock


\bibitem[Perozzi et~al\mbox{.}(2014)]%
        {perozzi2014deepwalk}
\bibfield{author}{\bibinfo{person}{B. Perozzi}, \bibinfo{person}{R. Al-Rfou},
  {and} \bibinfo{person}{S. Skiena}.} \bibinfo{year}{2014}\natexlab{}.
\newblock \showarticletitle{Deepwalk: Online learning of social
  representations}. In \bibinfo{booktitle}{\emph{KDD}}.
\newblock


\bibitem[Rahmattalabi et~al\mbox{.}(2021)]%
        {DBLP:conf/aaai/RahmattalabiJLV21}
\bibfield{author}{\bibinfo{person}{Aida Rahmattalabi}, \bibinfo{person}{Shahin
  Jabbari}, \bibinfo{person}{Himabindu Lakkaraju}, \bibinfo{person}{Phebe
  Vayanos}, \bibinfo{person}{Max Izenberg}, \bibinfo{person}{Ryan Brown},
  \bibinfo{person}{Eric Rice}, {and} \bibinfo{person}{Milind Tambe}.}
  \bibinfo{year}{2021}\natexlab{}.
\newblock \showarticletitle{Fair Influence Maximization: a Welfare Optimization
  Approach}. In \bibinfo{booktitle}{\emph{AAAI}}.
\newblock


\bibitem[Saito et~al\mbox{.}(2009)]%
        {Saito2009LearningCI}
\bibfield{author}{\bibinfo{person}{K. Saito}, \bibinfo{person}{Masahiro
  Kimura}, \bibinfo{person}{K. Ohara}, {and} \bibinfo{person}{H. Motoda}.}
  \bibinfo{year}{2009}\natexlab{}.
\newblock \showarticletitle{Learning Continuous-Time Information Diffusion
  Model for Social Behavioral Data Analysis}. In
  \bibinfo{booktitle}{\emph{ACML}}.
\newblock


\bibitem[Stoica et~al\mbox{.}(2020)]%
        {DBLP:conf/www/StoicaHC20}
\bibfield{author}{\bibinfo{person}{Ana{-}Andreea Stoica},
  \bibinfo{person}{Jessy~Xinyi Han}, {and} \bibinfo{person}{Augustin
  Chaintreau}.} \bibinfo{year}{2020}\natexlab{}.
\newblock \showarticletitle{Seeding Network Influence in Biased Networks and
  the Benefits of Diversity}. In \bibinfo{booktitle}{\emph{WWW}}.
\newblock


\bibitem[Tsang et~al\mbox{.}(2019)]%
        {groupinfluencemax2019}
\bibfield{author}{\bibinfo{person}{Alan Tsang}, \bibinfo{person}{Bryan Wilder},
  \bibinfo{person}{Eric Rice}, \bibinfo{person}{Milind Tambe}, {and}
  \bibinfo{person}{Yair Zick}.} \bibinfo{year}{2019}\natexlab{}.
\newblock \showarticletitle{Group-Fairness in Influence Maximization}. In
  \bibinfo{booktitle}{\emph{IJCAI}}.
\newblock


\bibitem[Wu et~al\mbox{.}(2021)]%
        {wu2021fairness}
\bibfield{author}{\bibinfo{person}{Chuhan Wu}, \bibinfo{person}{Fangzhao Wu},
  \bibinfo{person}{Xiting Wang}, \bibinfo{person}{Yongfeng Huang}, {and}
  \bibinfo{person}{Xing Xie}.} \bibinfo{year}{2021}\natexlab{}.
\newblock \showarticletitle{Fairness-aware news recommendation with decomposed
  adversarial learning}. In \bibinfo{booktitle}{\emph{Proceedings of the AAAI
  Conference on Artificial Intelligence}}, Vol.~\bibinfo{volume}{35}.
  \bibinfo{pages}{4462--4469}.
\newblock


\bibitem[Zhang et~al\mbox{.}(2013)]%
        {weibopaper}
\bibfield{author}{\bibinfo{person}{Jing Zhang}, \bibinfo{person}{Biao Liu},
  \bibinfo{person}{Jie Tang}, \bibinfo{person}{Ting Chen}, {and}
  \bibinfo{person}{Juanzi Li}.} \bibinfo{year}{2013}\natexlab{}.
\newblock \showarticletitle{Social Influence Locality for Modeling Retweeting
  Behaviors}. In \bibinfo{booktitle}{\emph{IJCAI}}.
\newblock


\end{thebibliography}
